\documentclass[10pt,journal,twocolumn]{IEEEtran}
\newif\ifCLASSOPTIONromanappendices \CLASSOPTIONromanappendicestrue
\usepackage{geometry}
\usepackage{a0size}        
\usepackage{amssymb}      
\usepackage{multicol}
\usepackage[english]{babel}    
\usepackage{epsfig} 
\usepackage{bm}
\usepackage{amsfonts,color,amsthm,amsmath, lscape,graphics}
\usepackage{url}
\usepackage{algorithm}  
\usepackage{algpseudocode}

\usepackage[font=footnotesize]{caption}
\usepackage[font=footnotesize]{subcaption}
\usepackage{subcaption}
\usepackage{epstopdf}
\usepackage{algorithm} 
\usepackage{algpseudocode}
\usepackage{cite}

\definecolor{awesome}{rgb}{1.0, 0.13, 0.32}
\usepackage{fix2col}

\addto\captionsenglish{} 
\newcommand{\changeb}[1]{{\color{black}#1}}
\newcommand{\changer}[1]{{\color{black}#1}}
\usepackage{graphicx}  
\theoremstyle{plain}

\usepackage{multirow} 
\usepackage{relsize}  
\usepackage{bm}   
\usepackage{pifont}

\newcommand{\argmax}{\arg\!\max} 
\newcommand{\argmin}{\arg\!\min}
\DeclareMathOperator{\sgn}{sgn}
\begin{document}
\title{Generalized Approximate Message Passing for  Massive MIMO mmWave Channel Estimation with Laplacian Prior}

\author{ Faouzi~Bellili, \IEEEmembership{Member, IEEE}, Foad~Sohrabi, \IEEEmembership{Member, IEEE} and~Wei~Yu, \IEEEmembership{Fellow, IEEE}
\thanks{This paper has been presented in part at IEEE 19th International Workshop on Signal Processing Advances in Wireless Communications (SPAWC), Kalamata, Greece, June 2018 \cite{SPAWC2018}. This work is supported in part by an NSERC Postdoctoral Fellowship grant and in part by the Canada Research Chairs program. F. Bellili, F. Sohrabi, and W. Yu are with the Edward S. Rogers Sr. Department
of Electrical and Computer Engineering, University of Toronto, Toronto, ON M5S 3G4, Canada (e-mails: faouzi.bellili@utoronto.ca; fsohrabi@comm.utoronto.ca; weiyu@comm.utoronto.ca). 
} 
} 
\maketitle
\begin{abstract}  
This paper tackles the problem of millimeter-Wave (mmWave) channel estimation in massive MIMO communication systems.
 A new Bayes-optimal channel estimator is derived using recent advances in the approximate belief propagation (BP) Bayesian inference paradigm.   By leveraging the inherent sparsity  of the mmWave MIMO channel in the angular domain, we recast the underlying channel estimation problem into that of reconstructing a compressible signal from a set of noisy linear measurements. Then, the  generalized approximate message passing (GAMP) algorithm is used to find the entries of the unknown mmWave MIMO channel matrix. Unlike all the existing  works on the same topic, we model the angular-domain channel coefficients by Laplacian distributed random variables. Further, we establish the closed-form expressions for the various statistical quantities that need to be updated iteratively by GAMP. To render
the proposed algorithm fully automated,  we also develop an \changeb{expectation-maximization (EM) based} procedure that can be easily embedded within  GAMP's iteration loop in order to learn all the unknown  parameters of the underlying Bayesian inference problem. Computer simulations  show that the proposed combined EM-GAMP algorithm under a Laplacian prior exhibits  improvements both in terms of channel estimation accuracy, achievable rate, and computational complexity as compared to the Gaussian mixture prior that has been advocated in the recent literature. In addition, it is found that  the Laplacian prior speeds up the convergence time of GAMP over the entire signal-to-noise ratio (SNR) range. 
\end{abstract}

\section{Introduction}
\IEEEPARstart{M}{assive} multiple-input multiple-output (MIMO) technology in which transceivers are equipped 
with a large number (tens to hundreds)  of  antennas has recently  attracted  considerable  research  interest  both  in 
academia  and  industry. In fact, owing  to  the  unprecedented   spectral  and  energy 
efficiency gains it promises, massive MIMO  is  foreseen  to  be  a  key  component and to play a major role in future fifth-generation (5G)  wireless 
networks  \cite{5G_1,5G_2,5G_3}.  Moreover, the combination of massive MIMO, mmWave, and  small-cell  
geometries  is  a  symbiotic  convergence  of  technologies, recognized  to  be  the  next  wireless 
revolution \cite{5G_4} which enables achieving the approximate thousand-fold increase in capacity that will be 
needed in the  coming  decades.  A common and crucial assumption behind many of the promised benefits of massive MIMO mmWave technology, however, is that the receiver and/or the transmitter are provided with good-quality channel state information (CSI),  which in practice has to be estimated using short-length pilot sequences.

As a matter of fact, the task of acquiring acceptable-quality CSI    is much more challenging for massive MIMO systems than it is for traditional MIMO configurations \cite{massive_MIMO_problem}. Indeed,  the direct application of conventional per-antenna channel estimation schemes leads to a prohibitive overhead.  This stems from  the need to transmit long training sequences to be able to accurately estimate all the entries of the large-size channel matrix.

Fortunately, unlike the ultra-high frequency (UHF) or sub-3GHz band, many recent channel measurement/sounding campaigns have confirmed that, due  to the lack of scattering  in mmWave bands, the signal propagates from the transmitter to the receiver through a small number of path clusters  \cite{mmwave_modeling_1,mmwave_modeling_2,mmwave_modeling_3}. This has triggered a surge of interest --- fueled by   recent progresses in compressed sensing (CS) theory --- in harnessing  the sparsity of mmWave channels to devise accurate estimators that require acceptable-size training sequences \cite{SP_overview}.    In this context, a plethora of massive MIMO mmWave channel estimators have been  introduced, over  the recent few years, by capitalizing mainly on the angular-domain sparsity. However, the vast majority of the proposed works  capture such sparsity by simply quantizing the beam domain. More specifically, a fixed sampling grid is selected to serve as a possible set for all candidate   values of the angles of departure (AoDs) and angles  of arrival (AoAs) pertaining to the different paths (see \cite{fixed_grid_0,fixed_grid_2} and references therein). Then, by constraining all true (unknown) AoDs/AoAs to lie exactly on the selected grid, the channel estimation problem is recast as a sparse reconstruction problem. Its  sensing matrix is  an  overcomplete dictionary that can be easily constructed by evaluating the transmit/receive array response vectors at  all the points of the postulated sampling grid. The sparsity of the \changeb{unknown vector,} which contains the path gain coefficients,  follows from the observation that most of the sampled AoD/AoA pairs do not correspond to any physical path\footnote{A sampled AoD/AoA pair that does not correspond to any physical path is associated with a zero coefficient with unknown location in $\mathbf{x}$. The idea of building an overcomplete dictionary via angular-domain quantization was pioneered in \cite{Willsky} within the framework of multisource DOA estimation under flat-fading channels.}. 

Unfortunately, the above techniques suffer from the inevitable off-grid problem which arises in practical situations where some of the true AoDs and/or AoAs do not lie on the sampling grid. For accurate estimation, it is therefore compulsory to use densely sampled grid since it reduces the gap between the true parameters and their nearest points on the grid. However, the cost of a dense grid sampling is the excessive increase in computational complexity.  

 From an algorithmic point of view, another common theme underlying almost all the existing methods is the use of sparsity-inducing mixed-norm optimization criteria, such as the basis pursuit \cite{basis_pursuit} (and its variants) or the LASSO \cite{lasso}, for reconstruction at the receiver. Although the existing convex optimization-based reconstruction algorithms  perform  well in practice, their complexity does not scale appropriately with the problem dimensions. Specifically, it increases very rapidly with the size of the transmit/receive antenna arrays as we operate in the massive MIMO regime.  
   To alleviate the computational burden,  popular iterative hard/soft thresholding (IHT/IST)  algorithms \cite{IST,IHT} can been envisaged here at the cost, however, of worse reconstruction performance. 
  
In this paper, we capitalize on the generalized approximate message-passing (GAMP) algorithm\footnote{Note here that the Bayesian MMSE version of the original AMP algorithm \cite{chen2018sparse} could also be used here and we expect  both algorithms to lead to similar results.} \cite{GAMP} to estimate the massive MIMO mmWave channel. Unlike the original AMP algorithm, pioneered in \cite{AMP_original}, GAMP is able to accommodate any priori distribution on the entries of the  unknown vector and applies to both linear and nonlinear observation models. GAMP has the reconstruction power of  mixed-norm optimization approaches and entails almost the same complexity of IST/IHT algorithms making it very attractive for large-dimensional estimation problems encountered in massive MIMO systems. However, it is known that GAMP  diverges with even mildly ill-conditioned sensing matrices  \cite{AMP_divergence_1,AMP_divergence_2,AMP_divergence_3}, like the overcomplete dictionaries resulting from beam-domain quantization. Similar to \cite{AMP_mmWave, AMP_multiuser, sparse_way}, to circumvent this problem, we rely on the so-called \textit{virtual} or \textit{canonical} channel model \cite{canonical} which amounts to expanding the physical multipath channel in terms of multidimensional Fourier basis functions that are easily implemented by discrete Fourier transform (DFT) matrices. This idea has been recently used together with the sparse message passing paradigm in \cite{Ref201,Ref202} within the context of massive MIMO mmWave channel estimation.     

The use of approximate message-passing algorithms has already found application in many different fields. But its application  to the problem of estimating  massive MIMO multipath sparse  channels  has appeared only recently in \cite{AMP_mmWave} and \cite{AMP_multiuser} for the case of single- and multi-user communications\footnote{We also refer the reader to \cite{Ref101, Ref102} for other  interesting recent works that apply the idea of Gaussian message passing to the detection problem in MIMO communications.}, respectively. In both  works, however,
GAMP is used in conjunction with  a Gaussian mixture (GM) prior on the channel coefficients.    
This paper makes an observation that  the use of a Laplacian prior on the beam-domain coefficients of massive MIMO mmWave channels can lead to better reconstruction performance while speeding up the convergence of GAMP at the same time.
  The use of a Laplacian prior in our work is, in part, motivated by its wide adoption in Bayesian image processing. There, it has been shown that the approximately sparse discrete cosine transform (DCT) coefficients of natural images are  better modeled by a Laplace distribution \cite{Laplace_image}. Owing to the apparent analogy between the  sparsity of DCT coefficients in image processing and the sparsity of mmWave MIMO channels in the DFT basis, it is expected that a Laplacian distribution will lead to enhanced reconstruction performance. In this paper, we establish in closed-from expressions all the statistical quantities required by GAMP under a Laplacian prior. Moreover,  we devise a simple approach that learns all the necessary parameters using the expectation-maximization (EM) principle. The proposed EM-based approach comes at almost no additional  cost since all the statistical quantities it requires are provided as  by-products of GAMP while trying to reconstruct the unknown channel itself. Exhaustive Monte-Carlo simulations  show  that a Laplacian prior leads indeed to large performance gains especially in the harsh conditions of low SNR and/or reduced number of observations. 
 
We structure the rest of this paper as follows. In section II, we present the system model.
 In section III, we recast the massive MIMO mmWave channel estimation problem into a sparse reconstruction problem and derive the GAM-based algorithm under a Laplacian prior. In Section IV,  we devise the EM-based approach that learns the parameters of the underlying estimation problem. In Section V, we discuss the simulation results of the algorithm. Finally, we draw out some concluding remarks in Section VI.

The common notations used in this paper are as follows. 
 Lower- and upper-case bold fonts, $\mathbf{x}$ and $\mathbf{X}$, are used denote vectors and matrices, respectively. Upper-case calligraphic font, $\mathcal{X}$ and $\bm{\mathcal{X}}$, is used to denote single and multivariate random variables, respectively. The $n$th column  of  $\mathbf{X}$ is denoted as $[\mathbf{X}]_{:,n}$, its $(m,n)$th entry is denoted as $\mathbf{X}_{mn}$, and the $n$th element of $\mathbf{x}$ is denoted as $x_n$.
  $\mathbf{I}_N$ stands for the $N\times N$ identity matrix, $\textrm{vec}(\mathbf{X})$ stacks the columns of $\mathbf{X}$ one below the other, and unvec$(.)$ is the associated inverse operator. The shorthand notation $\mathbf{x}\sim\mathcal{N}(\mathbf{m},\mathbf{R})$ means that the vector $\mathbf{x}$ follows a Gaussian distribution with mean $\mathbf{m}$ and auto-covariance matrix $\mathbf{R}$. Moreover, $\{.\}^\textsf{T}$ and $\{.\}^\textsf{H}$ stand for the transpose and Hermitian (transpose conjugate) operators, respectively. In addition, $|.|$ and $\|.\|$ stand for the modulus and Euclidean norm, respectively. Given any complex number,  $\Re\{.\}$, $\Im\{.\}$, and $\{.\}^*$ return its real part, imaginary part, and complex conjugate, respectively. The Kronecker  function and product are denoted as $\delta_{m,n}$ and $\otimes$, respectively.  We also denote the probability distribution function (pdf) of single and multivariate random variables (RVs) by $p_{\mathcal{X}}(x)$ and $p_{\bm{\mathcal{X}}}(\mathbf{x})$, respectively. The statistical expectation is denoted as $\mathbb{E}\{.\}$, $j$ is the imaginary unit (i.e., $j^{2}=-1$), and the notation $\triangleq$ is used for definitions.

\section{System Model}\label{section_2}
Consider a  massive MIMO mmWave communication system wherein the transmitter and the receiver  are equipped with $M_t$ and $M_r$ antenna branches, respectively. At  successive discrete time instants $k$, $k=0,1,\ldots,K-1$, the $i$th transmit antenna element sends a training (i.e., pilot) symbol $b_i(k)$. The entire $M_t-$dimensional signal broadcasted by the transmit antenna array  is denoted as $\mathbf{b}(k)\triangleq[b_1(k), b_2(k), \ldots, b_{M_t}(k)]^{\textsf{T}}$. This paper assumes that the elements of the pilot sequence are generated randomly from independent and identically distributed (i.i.d.) complex Gaussian distribution with zero mean and variance $1/\sqrt{M_t}$.   
 Under perfect carrier and timing recovery and assuming block fading, the $M_r-$dimensional received signals, denoted hereafter as $\mathbf{y}(k)$ for $k=0,1,\ldots,K-1$, can be modeled as follows: 
\begin{eqnarray}\label{system_model_general}
\!\!\!\!\mathbf{y}(k)&=&\mathbf{H}\mathbf{b}(k)~+~\mathbf{w}(k),
\end{eqnarray}  
where $\mathbf{H}\in \mathbb{C}^{M_r\times M_t}$ is the baseband channel impulse response, assumed to remain constant over the entire observation window, and $\mathbf{w}(k) \triangleq [w_1(k),w_2(k),\ldots, w_{M_r}(k)]^{\textsf{T}}$ is the additive noise vector at \changeb{discrete} time instant $k$. The noise components, $w_i(k)$, are modeled by circular complex Gaussian random variables with zero mean and variance $2\sigma_w^2$ and they are assumed to be temporally and spatially white, i.e., $\mathbb{E}\{w_i(k)w_{i'}(k')^*\} = 2\sigma_w^2\delta_{i,i'}\delta_{k,k'}$. 

It is worth mentioning here that the linear model in (\ref{system_model_general}) is valid only when a high precision ADC is used for each  antenna. In practice, the ADCs  have finite precision especially in mmWave massive MIMO communications  where the received data is coarsely quantized \cite{AMP_mmWave}. Quantization effects are not considered in this paper since our main goal is to investigate the impact of the prior distribution on the estimation of the sparse beam-domain channel coefficients. We show that the Laplacian  prior is the right model for beam-domain mmWave channel coefficients.  Extending the results of this work to  more practical architectures that use a dedicated RF chain with a low-precision ADC for each antenna is a promising future research direction. Indeed, even in the presence of finite-resolution ADCs, it is still expected that the Laplacian prior will be the right model  for the beam-domain coefficients of the mmWave massive MIMO channels. Another attractive hardware architecture for massive MIMO mmWave systems suggests to use a reduced number of RF chains (as compared to the number of antennas) but with full-precision ADCs, known as the hybrid architecture \cite{el2014spatially, zhang2005variable}.  
 In this respect, we refer the reader to  \cite{fixed_grid_2,zhou2017low, hu2018super} and references therein for recent works on the problem of channel estimation for massive MIMO mmWave systems with the hybrid architecture.  In particular, by evaluating the transmit and receive array steering vectors on a uniform grid of possible AoAs/AoDs, the authors of \cite{fixed_grid_2} recast the problem of channel estimation in massive MIMO mmWave systems with hybrid architectures as a compressed sensing problem. The unknown non-zero entries in the \textit{exactly} sparse vector, $\mathbf{x}$, of the resulting linear model are the gain coefficients of the individual physical paths. A key step to extending the results of \changeb{our} work to the hybrid architectures amounts to finding the appropriate linear model that has the beam-domain coefficients as the entries of the unknown sparse vector $\mathbf{x}$. Moreover, since GAMP applies to both linear and non-linear models, our results can potentially be further extended to hybrid mmWave architectures with finite-precision ADCs (i.e., with coarsely quantized observations) by relying on similar arguments in \cite{wen2016bayes}.   
 
The goal is to estimate the channel matrix,  $\mathbf{H}$, in (\ref{system_model_general}) given the set of observation vectors $\{\mathbf{y}(k)\}_{k=0}^{K-1}$.
In conventional sub-3GHz MIMO communications,
 the passing wave tends to hit the receiving antenna array from almost all directions. In this case, every single entry of   $\mathbf{H}$ is the aggregate contribution  of a large number of propagation paths and, hence, reasonably modeled by a complex Gaussian RV.    
  In higher-frequency mmWave bands, however,  
  electromagnetic waves 
   behave quite differently and radio signals propagate along very few path clusters each containing a small number of sub-paths with small angular \changeb{spreads}. As a result, the mmWave channel is inherently sparse in the angular domain if expressed in  suitable DFT bases \cite{sparse_way}.

To see this,  assume that the antenna elements form a  uniform linear array (ULA) configuration both at  the transmitter and receiver sides.  Assume also that there are $P$ clusters and that within each
     $p$th cluster ($p=1,2,\ldots, P$) there are $Q_p$ sub-paths. Moreover, each
       $q$th sub-path within the $p$th cluster ($q=1,2,\ldots, Q_p$) has an angle of departure $\phi_{p,q}$ and impinges on the receive antenna array from an angle of arrival $\theta_{p,q}$. By the superposition principle, the resulting baseband MIMO channel matrix is given by (see \cite{Ref202} and references therein): 
\begin{eqnarray}\label{channel_model}
\mathbf{H}&=& \sum_{p=1}^{P}\sum_{q=1}^{Q_p}\alpha_{p,q}\mathbf{a}_r(\Omega^r_{p,q})\mathbf{a}_t(\Omega^t_{p,q})^{\textsf{H}},
\end{eqnarray}
where $\alpha_{p,q}$ is the gain of the $q^{th}$  sub-path  within the $p^{th}$ cluster, $\Omega^t_{p,q} = \cos(\phi_{p,q})$  and $\Omega^r_{p,q}=\cos(\theta_{p,q})$  are its  directional cosine  with respect to the transmit and receive antenna arrays, respectively. Moreover, $\mathbf{a}_t(\Omega)$ and $\mathbf{a}_r(\Omega)$ are, respectively, the transmit and receive array response vectors which are explicitly given by \cite{Van_Trees}:
\begin{eqnarray}
\mathbf{a}_t(\Omega)&=& \big[1, e^{-j\pi\Omega}, e^{-j2\pi\Omega}, \ldots ,e^{-j(M_t-1)\pi\Omega}\big]^{\textsf{T}},\label{steering_t}\\
\mathbf{a}_r(\Omega)&=& \big[1,e^{-j\pi\Omega},e^{-j2\pi\Omega},\ldots,e^{-j(M_r-1)\pi\Omega}\big]^{\textsf{T}}\label{steering_r},
\end{eqnarray}  
where it is implicitly assumed that the antenna elements are separated by half the wavelength.

 The actual channel matrix as expressed in (\ref{channel_model}) is not visibly sparse. Expressing it in the angular domain, however, reveals that it has indeed  very few dominant entries. To see this, consider the following angular-domain representation of $\mathbf{H}$:
\begin{eqnarray} \label{angular_reprsentation}
\widetilde{\mathbf{H}} &=& \mathbf{U}_r^{\textsf{H}}~\!\mathbf{H}~\!\mathbf{U}_t,
\end{eqnarray}  
where $\mathbf{U}_r$  and $\mathbf{U}_t$ are the $M_r\!\times\! M_r$ and $M_t\!\times\!M_t$ spatial unitary DFT matrices. Plugging (\ref{angular_reprsentation}) back into (\ref{channel_model}) and rearranging the terms, it follows that:
\begin{eqnarray}\label{channel_model_angular}
\!\!\widetilde{\mathbf{H}}&=& \sum_{p=1}^{P}\sum_{q=1}^{Q_p}\alpha_{p,q}\Big[\mathbf{U}_r^{\textsf{H}}\,\mathbf{a}_r(\Omega^r_{p,q})\Big]\!\Big[\mathbf{U}_t^{\textsf{H}}\,\mathbf{a}_t(\Omega^t_{p,q})\Big]^{\textsf{H}}\!\!. 
\end{eqnarray}

Clearly, in (\ref{channel_model_angular}), the columns of $\mathbf{U}_r$  and $\mathbf{U}_t$  act as  receive and transmit beamforming vectors, respectively, capturing how much energy is present along their associated transmit/receive \changeb{beams.} In fact, by considering the $m^{th}$ component of the vector $\mathbf{v}_{p,q}^r~\!\triangleq~\!\mathbf{U}_r^{\textsf{H}}\mathbf{a}_r(\Omega^r_{p,q})$:
\begin{eqnarray}
\mathbf{v}_{p,q}^r[m] &=& [\mathbf{U}_{r}]_{:,m}^{\textsf{H}}\!~\mathbf{a}_r(\Omega^r_{p,q}),~~m=1,2,\ldots, M_r,
\end{eqnarray}
it can be shown that its magnitude is explicitly  expressed as follows:
\begin{eqnarray}
~\Big|\mathbf{v}_{p,q}^r[m]\Big| ~=~ \frac{1}{\sqrt{M_r}}\left|\frac{\sin\Big(\pi\big[m-1-\frac{M_r}{2}\Omega^r_{p,q}\big]\Big)}{\sin\Big(\frac{\pi}{M_r}\big[m-1-\frac{M_r}{2}\Omega^r_{p,q}\big]\Big)}\right|.
\end{eqnarray} 

It is clear that $\big|\mathbf{v}_{p,q}^r[m]\big|$ is maximal for $m_0$ verifying:
\begin{eqnarray}\label{maximal_energy}\label{AoA_condition}
\Big|\Omega^r_{p,q}-\frac{2(m_0-1)}{M_r}\Big|<\frac{2}{M_r}.
\end{eqnarray}
This is in line with a standard result in array signal processing which states that each of the receive beamforming vectors $[\mathbf{U}_{r}]_{:,m}$, for  $m=1,2,\ldots,M_r$, has a main lobe centered around $(m-1)/M_r\Delta_r$ with beamwidth $1/(M_r\Delta_r)$ where $\Delta_r$ is the inter-antenna separation normalized by the wavelength (in our case $\Delta_r = 1/2$). 
 Therefore, a given subpath with receive directional cosine $\Omega^r_{p,q}$  has almost all of  its energy along one particular vector $[\mathbf{U}_{r}]_{:,m_0}$  \big(see (\ref{maximal_energy})\big) and very little along all the others. If the angular spread, $\sigma_{\theta}^r$, is small (typically $\sigma_{\theta}^r<1/M_r\Delta_r$), then all the sub-paths belonging to the same cluster have most of their energy along the same beamforming vector.  By the same virtue, they also have most of their energy concentrated along one particular transmit beamforming vector $[\mathbf{U}_{t}]_{:,n_0}$ for  some $n_0 \in \{1,2,\ldots, M_t-1\}$ that satisfies: 
\begin{eqnarray}\label{AoD_condition}  
\Big|\Omega^t_{p,q}-\frac{2(n_0-1)}{M_t}\Big|<\frac{2}{M_t}.
\end{eqnarray}

In this way, the only few dominant entries, $\widetilde{\mathbf{H}}_{m_0n_0}$, of $\widetilde{\mathbf{H}}$  are those for which there is a cluster with mean AoA and mean AoD that verify (\ref{AoA_condition}) and  (\ref{AoD_condition}) at the same time. All the remaining  entries  have relatively small magnitude; they are not identically zero as  they capture small contributions from all the clusters due to spectral leakage phenomena.    

\section{Proposed mmWave channel estimation algorithm}
\subsection{Problem formulation}
To harness the angular domain sparsity of the channel, we propose to apply the DFT precoder, $\mathbf{U}_t^{\textsf{H}}$, to the  training sequences before transmission, i.e., the transmit array sends $\mathbf{U}_t\mathbf{b}(k)$ at discrete time instants $k$, $k=0,1,\ldots, K-1$.  We also combine the corresponding  received noisy vector using the DFT combiner, $\mathbf{U}_r$, such that:
\begin{subequations}
\begin{eqnarray}\label{system_model_combining}
\widetilde{\mathbf{y}}(k)&=&\mathbf{U}_r^{\textsf{H}}\mathbf{H}\mathbf{U}_t\mathbf{b}(k)~+~\mathbf{U}_r^{\textsf{H}}\mathbf{w}(k),\\
\label{system_model_combining11}&=&\widetilde{\mathbf{H}}\,\mathbf{b}(k) ~+~\widetilde{\mathbf{w}}(k),
\end{eqnarray}     
\end{subequations}
where $\widetilde{\mathbf{w}}(k) \triangleq \mathbf{U}_r^{\textsf{H}}\mathbf{w}(k)$ is the resulting combined noise which has exactly the same statistics as $\mathbf{w}(k)$ since the matrix $\mathbf{U}_r$ is unitary. By stacking all the received vectors in a single matrix $\widetilde{\mathbf{Y}} = [\widetilde{\mathbf{y}}(0),\widetilde{\mathbf{y}}(1),\ldots,\widetilde{\mathbf{y}}(K-1)]$, we obtain:
\begin{eqnarray}\label{system_model}
\widetilde{\mathbf{Y}} &=& \widetilde{\mathbf{H}}\mathbf{B}~+~\widetilde{\mathbf{W}},
\end{eqnarray}
with the matrices $\mathbf{B}$ and $\widetilde{\mathbf{W}}$ being constructed in the same way as $\mathbf{Y}$, i.e., $\mathbf{B} = [\mathbf{b}(0),\mathbf{b}(1),\ldots,\mathbf{b}(K-1)]$ and $\widetilde{\mathbf{W}} = [\widetilde{\mathbf{w}}(0),\widetilde{\mathbf{w}}(1),\ldots,\widetilde{\mathbf{w}}(K-1)]$. Now,  vectorizing (\ref{system_model}) yields:
\begin{eqnarray}\label{system_model_vec}
\textrm{vec}\big(\widetilde{\mathbf{Y}}\big) &=& (\mathbf{B}^{\textsf{T}}\otimes \mathbf{I}_{M_r})\textrm{vec}\big(\widetilde{\mathbf{H}}\big)~+~\textrm{vec}\big(\widetilde{\mathbf{W}}\big).
\end{eqnarray} 
By defining $\widetilde{\mathbf{y}}\triangleq \textrm{vec}(\widetilde{\mathbf{Y}})\in\mathbb{C}^{M_r K}$,  $\widetilde{\mathbf{x}}\triangleq \textrm{vec}(\widetilde{\mathbf{H}})\in\mathbb{C}^{M_rM_t}$, $\widetilde{\mathbf{w}}\triangleq \textrm{vec}(\widetilde{\mathbf{W}})\in\mathbb{C}^{M_rK}$, and $\widetilde{\mathbf{A}}\triangleq \mathbf{B}^{\textsf{T}}\otimes \mathbf{I}_{M_r}\in\mathbb{C}^{M_r K\times M_rM_t}$, (\ref{system_model_vec}) is equivalently rewritten  as follows:
\begin{eqnarray}\label{system_model_vec_generic_0}
\widetilde{\mathbf{y}} &=& \widetilde{\mathbf{A}}\widetilde{\mathbf{x}}~+~\widetilde{\mathbf{w}}.
\end{eqnarray} 
Recall here that the vector $\widetilde{\mathbf{x}}$ is approximately sparse due to the approximate sparsity of $\widetilde{\mathbf{H}}$. This paper captures the underlying sparsity by a Laplacian distribution.  Since the Laplacian distribution is defined for real-valued RVs only, we transform the complex  model in (\ref{system_model_vec_generic_0}) to the following equivalent real model:
\begin{eqnarray}\label{transfromed_model}
     \underbrace{\begin{bmatrix}
       \Re\{\widetilde{\mathbf{y}}\}\, \\[0.4em]      
      \Im\{\widetilde{\mathbf{y}}\}
     \end{bmatrix}}_{\substack{\\\mathlarger{\triangleq}\\\mathlarger{\mathbf{y}}}}
     &\!\!=\!\!& 
     \underbrace{\begin{bmatrix}
       \Re\{\widetilde{\mathbf{A}}\} & -\Im\{\widetilde{\mathbf{A}}\}          \\[0.4em]
       \Im\{\widetilde{\mathbf{A}}\}          & \Re\{\widetilde{\mathbf{A}}\}
     \end{bmatrix}}_{\substack{\\\mathlarger{\triangleq}\\\mathlarger{\mathbf{A}}}} 
     \underbrace{\begin{bmatrix}
       \Re\{\widetilde{\mathbf{x}}\} \\[0.4em]      
      \Im\{\widetilde{\mathbf{x}}\}
     \end{bmatrix}}_{\substack{\\\mathlarger{\triangleq}\\\mathlarger{\mathbf{x}}}}
     ~+~
     \underbrace{\begin{bmatrix}
       \Re\{\widetilde{\mathbf{w}}\} \\[0.4em]      
      \Im\{\widetilde{\mathbf{w}}\}
     \end{bmatrix}}_{\substack{\\\mathlarger{\triangleq}\\\mathlarger{\mathbf{w}}}}.\nonumber\\
\end{eqnarray} 
  
 We recognize in (\ref{transfromed_model}) the well-known  \textit{inverse problem} in signal  processing research practices:  reconstruct a (approximately) sparse vector from the fewest possible number of noisy linear  observations: 
 \begin{eqnarray}\label{system_model_vec_generic}
\mathbf{y} &=& \mathbf{A}\mathbf{x}~+~\mathbf{w},
\end{eqnarray} 
in which by defining $M=2M_r K$ and $N=2M_rM_t$, we have  $\mathbf{y}\in\mathbb{R}^{N}$, $\mathbf{A}\in\mathbb{C}^{M\times N}$, and $\mathbf{x}\in\mathbb{C}^{N}$. 
    Once this is done, an estimate of the original channel matrix is readily obtained from (\ref{channel_model_angular}) as follows:
\begin{eqnarray}
\widehat{\mathbf{H}}~=~\mathbf{U}_r\widehat{\widetilde{\mathbf{H}}}\mathbf{U}_t^{\textsf{H}}, 
\end{eqnarray}   
where $\widehat{\widetilde{\mathbf{H}}} =\textrm{unvec}\big(\widehat{\widetilde{\mathbf{x}}}\big)$  and $\widehat{\widetilde{\mathbf{x}}}$ is an estimate of  $\widetilde{\mathbf{x}}$ which is easily obtained from the reconstructed vector $\widehat{\mathbf{x}}$.

\subsection{Modeling the Angular-Domain Coefficients of mmWave Channels:}
In our quest for finding  the beam-domain channel coefficients in $\mathbf{x}$, we follow the Bayesian approach which  requires an appropriate model for the  prior distribution of the $x_n$'s. For rich-scattering environments, an accurate and widely used statistical model for the actual channel coefficients is the Gaussian model. In other words, the various entries of the channel matrix $\mathbf{H}$ involved in (\ref{system_model_general}) are  assumed to follow a scaled normal distribution:
\begin{eqnarray}\label{scaled_distribution_Gaussian}
p_{\mathcal{X}}(x;\beta) = \mathcal{N}(x;0,\beta),
\end{eqnarray}
in which $\beta$ is the large-scale fading coefficient that accounts for  the combined effects of shadowing and path loss. In this case, the entries of $\widetilde{\mathbf{H}}$ in (\ref{system_model}) or equivalently the elements of the unknown vector $\mathbf{x}$ in (\ref{system_model_vec_generic})  
are also Gaussian-distributed
since $\mathbf{U}_r$ and $\mathbf{U}_t$ are unitary matrices.

   In mmWave communications, however, the entries of $\mathbf{H}$ cannot be approximated by a Gaussian distribution due to the lack of scattering. Hence, a more appropriate statistical model for the angular-domain channel coefficients needs to be specified.    
 This problem has been  addressed  in  two recent works \cite{AMP_mmWave} and \cite{AMP_multiuser} for single- and multi-user communications, respectively, and  both of these works
 propose to approximate the unknown prior distribution by a Gaussian mixture (GM) model of order $L$, i.e.:
\begin{eqnarray}\label{GM_model}
p_{\mathcal{X}}(x;\bm{\alpha})&=&\sum_{l=1}^L\omega_l\mathcal{N}(x;\eta_l, \nu_l),
\end{eqnarray}
where $\{\omega_l\}_{l=1}^{L}$ are the normalized mixing coefficients in the postulated GM model that is parameterized by the vector $\bm{\alpha}\triangleq [\omega_1,\ldots\omega_L,\eta_1,\ldots,\eta_L, \nu_1,\ldots,\nu_L]^{\textsf{T}}$.  In \cite{AMP_mmWave,AMP_multiuser}, the  components of $\bm{\alpha}$ are estimated by the expectation-maximization procedure for the GM model introduced recently in \cite{Schniter_EM_GM}.

In this paper, we propose to model the angular-domain coefficients of the underlying mmWave \changeb{massive} MIMO channel by a zero-mean Laplacian distribution with scale parameter $b$: 
\begin{eqnarray}\label{Laplace_distribution}
p_{\mathcal{X}}(x;b) &=& \frac{1}{2b}\mathlarger{e^{-\frac{|x|}{b}}}.
\end{eqnarray} 
As mentioned previously, our choice is motivated by the widespread use of the Laplacian distribution to capture the sparsity of DCT coefficients of natural images \cite{Laplace_image}.
 Moreover, it can be shown that  MAP-based estimation of sparse signals with Laplacian prior is equivalent to the  regularized $l_1-$norm minimization problem which is known to promote sparsity. Indeed, as explained in \cite{Laplace_sparsity,Laplace_sparsity_1}, the Laplacian prior enforces the sparsity constraint more heavily by distributing the posterior mass more on the axes so that signal coefficients close to zero are preferred.
 
\changeb{The} superiority of the Laplacian prior over the GM prior is intuitively expected. Indeed, both priors are special cases of the so-called generalized Gaussian scale  model (GSM) which yields a richer class of priors. To see this, consider a zero-mean RV $\mathcal{X}$ which is given by:
\begin{eqnarray}
\mathcal{X}&=&\sqrt{\mathcal{Z}}\mathcal{U}~~~~~~\textrm{with}~~~~~~~~ \mathcal{U}\sim\mathcal{N}(u;0,1),
\end{eqnarray}
and ${\mathcal{Z}}$ being a positive random variable which is statistically independent from  $U$. Hence:
 \begin{eqnarray}
 p_{\mathcal{X}}(x)&=&\int_{-\infty}^{+\infty}p_{\mathcal{Z}}(z)p_{\mathcal{X}|\mathcal{Z}}(x|z)\textsf{d}z\\
\label{GSM_Rev_2} &=&\int_{-\infty}^{+\infty}p_{\mathcal{Z}}(z)\mathcal{N}(x,0,z)\textsf{d}z
 \end{eqnarray}    
 By closely inspecting (\ref{GSM_Rev_2}), it can be shown that when ${\mathcal{Z}}$ is a discrete-valued RV with finite support $p_{\mathcal{X}}(x)$ reduces to a \textit{finite} mixture of Gaussian densities. Then, a more elaborate model would be to use an \textit{uncountably infinite} mixture (i.e., integral) of Gaussian densities by considering a continuous-valued RV ${\mathcal{Z}}$. One particular choice that is appealing from the computational point of view is to use an exponential distribution for ${\mathcal{Z}}$. By doing so,  one actually recovers the Laplacian prior that we are using in our paper. It becomes clear then that the Laplacian prior would  lead to better performance than the finite GM model unless one is willing to use a GM model that mixes a large number of normal distributions whose parameters become very difficult to learn in practice. Indeed, by using a Laplacian prior, one is actually using a GM model which involves the  sum of uncountably infinite number of Gaussian distributions while requiring to learn only one parameter. By learning this parameter,  we are implicitly learning the mixing coefficients along with the variances involved in the infinite-sum (i.e., integral) of the GSM model; instead of learning the individual parameters of the finite GM model as done in previous works.
 
We emphasize here the fact that, unlike the GM model in (\ref{GM_model}), the Laplacian distribution in (\ref{Laplace_distribution}) is parameterized by a single parameter $b$ which is itself unknown in advance. Later in this Section, we show how it can also be learned adaptively using the EM principle by exploiting the auxiliary outputs of GAMP. It is noteworthy, however,  that learning a single parameter (namely $b$) is statistically more efficient and entails much less computational complexity than learning  $3L$ parameters under the GM model.  Note as well that the large-scale fading coefficient is also assumed to be unknown and absorbed in the scale parameter $b$.

It is also noteworthy that the original AMP algorithm \cite{AMP_original} implicitly uses a Laplacian prior on the components of the unknown sparse vector $\mathbf{x}$.  However, the denoiser in the original AMP paper is not designed specifically for this Laplacian prior. Instead the denoiser design is based \changeb{on} the minimax criterion, which results in soft/hard  thresholding of the components of $\mathbf{x}$. It is the generalized AMP (GAMP) algorithm that offers a systematic way of taking the Laplacian (and actually any) prior into account during the denoising step, as will be explained in the next section.

Note also that in this paper we do not take into account the correlation between the angular-domain coefficients of mmWave massive MIMO channels. Indeed, the significant elements (in the angular domain) of  
mmWave massive MIMO channels usually appear in bursts due to the physical scattering structure. In order to capture that correlation, one needs to find an appropriate prior, $p_{\bm{\mathcal{X}}}(\mathbf{x})$, for the entire vector $\mathbf{x}$ and then estimate the latter using  \textit{vector} (instead of \textit{scalar}) GAMP. This requires, however, the inversion of multiple large-size matrices at every iteration on the top of learning  the entire covariance matrix of $\mathbf{x}$   using the EM algorithm instead of learning a single scale parameter as done in this paper.

\subsection{MmWave Channel Estimation using GAMP with Laplacian Prior }
GAMP algorithm applies loopy belief propagation on the bipartite graph obtained from (\ref{system_model_vec_generic}) under Gaussian approximations for the involved messages which become accurate in the large system limit. It falls under the Bayesian estimation framework wherein by assuming a prior distribution,
 $p_{\bm{\mathcal{X}}}(\mathbf{x})$ on $\mathbf{x}$
  one is interested  in finding the marginal posterior distributions $p_{\mathcal{X}_n|\bm{\mathcal{Y}}}(x_n|\mathbf{y})$. If possible, these could be used to perform  minimum mean square error (MMSE) or maximum \textit{a posteriori} estimation of each $x_n$ separately:
\begin{eqnarray}
\widehat{x}_{n}^{\textsc{MAP}} &=& \argmax_{x_n}p_{\mathcal{X}_n|\bm{\mathcal{Y}}}(x_n|\mathbf{y}),\\
\widehat{x}_{n}^{\textsc{MMSE}}&=&\argmin_{\widehat{x}_n}\mathbb{E}_{\mathcal{X}_n,\bm{\mathcal{Y}}}\big\{(x_n-\widehat{x}_n)^2\big\}\\
&=&\mathbb{E}_{\mathcal{X}_n|\bm{\mathcal{Y}}}\{x_n|\mathbf{y}\}.
\end{eqnarray}

Unfortunately, finding the true marginal distributions, $p_{\mathcal{X}_n|\bm{\mathcal{Y}}}(x_n|\mathbf{y})$, is analytically intractable  and  computationally prohibitive. To sidestep this problem, GAMP implements
 loopy belief propagation and relies on the central limit theorem (CLT) and quadratic approximations to solve the MMSE and MAP estimation problems, respectively.  Specifically, the sum-product and max-sum BP algorithms are used in the former and latter cases, respectively. In the sequel, we focus on  MMSE estimation wherein GAMP  approximates $p_{\mathcal{X}_n|\bm{\mathcal{Y}}}(x_n|\mathbf{y})$  by another tractable distribution that is progressively refined from one iteration to another.   In fact, assume that the components, $\{x_n\}_{n=1}^N$, of the unknown vector $\mathbf{x}$ are independent\footnote{Note here that the real and imaginary parts of each angular-domain channel coefficient can be dependent. In principle, this dependence can be captured by a bivariate Laplacian distribution. However, the latter does not enable one to find analytical expressions for the posterior mean and variance involved in Lines 15 and 16 of Algorithm 1, respectively.}
  and identically distributed (i.i.d.) according to a common \textit{prior} distribution, $p_{\mathcal{X}}(x;\bm{\alpha})$, which is parameterized by an unknown parameter vector $\bm{\alpha}$.
  After being linearly transformed to produce $\mathbf{z}\triangleq \mathbf{A}\mathbf{x}$, the latter is propagated through a probabilistic channel
\begin{eqnarray}
p_{\bm{\mathcal{Y}}|\bm{\mathcal{Z}}}\left(\mathbf{y}|\mathbf{z};\sigma_w^2\right)&\!\!=\!\!&\prod_{m=1}^Mp_{\mathcal{Y}_m|\mathcal{Z}_m}\left(y_m|z_m;\sigma_w^2\right). 
\end{eqnarray}

Although both $\bm{\alpha}$ and $\sigma_w^2$ need to be estimated as well (cf. Section IV) assume here and in the next subsection that they are perfectly known to the receiver and gather them in a single parameter vector    $\bm{\theta}\triangleq [\bm{\alpha},\sigma_w^2]$. Given knowledge of $\mathbf{y}$, $\mathbf{A}$, and $\bm{\theta}$, GAMP runs iteratively according to the algorithmic description provided in Algorithm~\ref{alg_GAMP}.    

\begin{algorithm}[t]v
  \caption{Sum-Product GAMP for MMSE estimation}
  \label{alg_GAMP}
  \begin{algorithmic}[1]\vskip 0.1cm
    \Require
      $\mathbf{A}\in \mathbb{R}^{M\times N}$; $\mathbf{y}\in\mathbb{R}^{M}$; $\bm{\alpha}$; $\sigma_w^2$, $p_{\mathcal{X}}(x;\bm{\alpha})$, precision tolerance ($\epsilon$), maximum number of iterations ($T_\textsc{max}$)\vskip 0.2cm
    \Ensure
      MMSE estimates $\{\widehat{x}_n\}_{n=1}^{N}$ for $\{x_n\}_{n=1}^{N}$   \\ \vskip 0.3cm  
     \textbf{Initialization}\\
      $t\gets 1$\\\vskip 0.1cm
           $\forall n:~\widehat{x}_n(t)\, =\, \int_{x}x\,p_{\mathcal{X}}(x;\bm{\alpha})\,\mathrm{d}x$\\\vskip 0.1cm
        $\forall n:~\mu^x_n(t) \,=\, \int_{x}|x-\widehat{x}_n(t)|^2p_{\mathcal{X}}(x;\bm{\alpha})\,\mathrm{d}x$\\\vskip 0.1cm
        $\forall m:~\widehat{s}_m(t-1)\,=\,0$\vskip 0.3cm
    \Repeat 
\State $\forall m:~\mu_m^p(t) ~=~ \sum_{n=1}^{N}|\mathbf{A}_{m,n}|^2\mu_n^x(t)$\vskip 0.1cm
\State $\forall m:~\widehat{p}_m(t)\, =\, \sum_{n=1}^{N}\mathbf{A}_{m,n}\widehat{x}_n(t)\,-\,\mu_m^p(t)\widehat{s}_m(t-1)$\vskip 0.1cm
\State $\forall m:~\mu_m^z(t) ~=~ \textsf{var}_{\mathcal{Z}_m\big|\bm{\mathcal{Y}}}\Big\{z_m|\mathbf{y}\,;\,\widehat{p}_m(t),\mu^p_m(t),\bm{\theta}\Big\}$\vskip 0.1cm
\State $\forall m:~\widehat{z}_m(t) ~=~ \mathbb{E}_{\mathcal{Z}_m|\bm{\mathcal{Y}}}\Big\{z_m\big|\mathbf{y}\,;\,\widehat{p}_m(t),\mu^p_m(t),\bm{\theta}\Big\}$\vskip 0.1cm
\State $\forall m:~\mu_m^s(t)\, =\, \frac{1}{\mu_m^p(t)}\left[1-\frac{\mu_m^z(t)}{\mu_m^p(t)}\right]$\vskip 0.1cm
\State $\forall m:~\widehat{s}_m(t) ~=~ \frac{1}{\mu_m^p(t)}\big[\widehat{z}_m(t)-\widehat{p}_m(t)\big]$\vskip 0.1cm
\State $\forall n:~\mu_n^r(t) ~=~ \left(\sum_{n=1}^{N}|\mathbf{A}_{m,n}|^2\mu_m^s(t)\right)^{-1}$\vskip 0.1cm
     \State $\forall n:~\widehat{r}_n(t)~ =~ \widehat{x}_n(t)~+~ \mu^r_n(t)\sum_{m=1}^{M}\mathbf{A}^{*}_{m,n}\widehat{s}_m(t)$\vskip 0.1cm
           \State $\forall n:~\widehat{x}_n(t+1) ~=~ \mathbb{E}_{\mathcal{X}_n|\bm{\mathcal{Y}}}\Big\{x_n\big|\mathbf{y}\,;\,\widehat{r}_n(t),\mu^r_n(t),\bm{\theta}\Big\}$\vskip 0.1cm
     \State  $\forall n:~\mu^x_n(t+1)~ =~ \textsf{var}_{\mathcal{X}_n|\bm{\mathcal{Y}}}\Big\{x_n\big|\mathbf{y}\,;\,\widehat{r}_n(t),\mu^r_n(t),\bm{\theta}\Big\}$\vskip 0.1cm
       \State  $t\gets t+1$\vskip 0.3cm 
    \Until{$\big|\!\big|\widehat{\mathbf{x}}(t+1)\,-\,\widehat{\mathbf{x}}(t)\big|\!\big|^2\leq\epsilon\,\big|\!\big|\,\widehat{\mathbf{x}}(t)\big|\!\big|^2$~~\textsf{or}~~ $t>T_\textsc{max}$}
  \end{algorithmic}
\end{algorithm} 
First, observe that the message propagated along the weak edge from the factor node $p(z_m|\mathbf{a}_m^{\textsf{T}}\mathbf{x})$ to the variable node $z_m$ is computed by integrating over all the variable nodes $\{x_n\}_n$ in $\mathbf{x}$ (here $\mathbf{a}_m^{\textsf{T}}$ is the $m$th row of $\mathbf{A}$). Therefore, using the CLT argument --- due to integration over a large number of variables ---  GAMP approximates this message by a Gaussian distribution with mean $\widehat{p}_m$ and variance $\mu^p_m$ (obtained from lines 7 and 8 of Algorithm 1), i.e., $\mathcal{N}(z_m;\widehat{p}_m,\mu^p_m)$. By taking the product of both incoming messages at node $z_m$, i.e., those outgoing from factor nodes $p(z_m|\mathbf{a}_m^{\textsf{T}}\mathbf{x})$ and $p_{\mathcal{Y}_m|\mathcal{Z}_m}(y_m|z_m;\bm{\theta})$, 
 the marginal posterior $p_{\mathcal{Z}_m|\bm{\mathcal{Y}}}(z_m|\mathbf{y};\bm{\theta})$ can be approximated by: 
\begin{eqnarray}\label{psoterior_prob_z}
\!\!\!\!\!\!\!\!p_{\mathcal{Z}_m|\bm{\mathcal{Y}}}\big(z_m|\mathbf{y};\widehat{p}_m,\mu^p_m,\bm{\theta}\big)&\!\!\!\!\!\!\!\!&\nonumber\\
&\!\!\!\!\!\!\!\!&\!\!\!\!\!\!\!\!\!\!\!\!\!\!\!\!\!\!\!\!\!\!\!\!\!\!\!\!\!\!\!\!=\frac{p_{\mathcal{Y}_m|\mathcal{Z}_m}(y_m|z_m;\bm{\theta})\mathcal{N}(z_m;\widehat{p}_m,\mu^p_m)}{\int_{z}p_{\bm{\mathcal{Y}}|\mathcal{Z}_m}(y_m|z_m;\bm{\theta})\mathcal{N}(z_m;\widehat{p}_m,\mu^p_m)}.
\end{eqnarray}

 It is seen that (\ref{psoterior_prob_z}) holds irrespectively of the prior distribution $p_{\mathcal{X}}(x;.)$, therefore, the posterior mean and variance (in lines 9 and 10) of $\mathcal{Z}_m$  under the Laplacian distribution are the same as for the GM model and their expressions are readily available from \cite{GAMP} as:
 \begin{eqnarray}
 \widehat{z}_m(t)&=& \frac{\mu^p_my_m~+~\sigma_w^2\widehat{p}_m}{\mu^p_m~+~\sigma_w^2},\\
 \mu_m^z(t) &=& \frac{\mu^p_m\sigma_w^2}{\mu^p_m~+~\sigma_w^2}
 \end{eqnarray} 
The quantities $\widehat{s}_m(t)$ and $\mu_m^s(t)$ updated in lines 12 and 11 of algorithm 1, respectively, are the equivalent of the Onsager correction term in the original AMP algorithm \cite{AMP_original}. They follow directly from the quadratic approximations involved in the general theory of GAMP.  

The message propagated along the strong edges from all the factor nodes $\{p(z_m|\mathbf{a}_m^{\textsf{T}}\mathbf{x})\}_m$ to the variable node $x_m$ is computed by integrating over all the \textit{other} variable nodes, $\{x_n'\}_{n'\neq n}$, in $\mathbf{x}$ and all the $z_m$'s in $\mathbf{z}= \mathbf{A}\mathbf{x}$. Therefore, using the CLT argument again, GAMP approximates this message by a Gaussian distribution with mean $\widehat{r}_m$ and variance $\mu^r_m$ (updated in lines 14 and 13 of Algorithm 1), i.e., $\mathcal{N}(x_m;\widehat{r}_m,\mu^r_m)$.  
Therefore, the marginal posterior  $p_{\mathcal{X}|\bm{\mathcal{Y}}}(x|\mathbf{y};\bm{\theta})$ at iteration $t$ can be approximated by taking the product of the aforementioned incoming Gaussian message, $\mathcal{N}(x_m;\widehat{r}_m,\mu^r_m)$, and any (common) prior distribution, $p_{\mathcal{X}}(x_n;\bm{\alpha})$, on the components  of $\mathbf{x}$:

\begin{eqnarray}\label{psoterior_prob}
\!\!\!\!\!\!\!\!\!\!p_{\mathcal{X}_n|\bm{\mathcal{Y}}}(x_n|\mathbf{y};\widehat{r}_n,\mu^r_n,\bm{\theta})&\!\!=\!\!&\frac{p_{\mathcal{X}}(x_n;\bm{\alpha})\mathcal{N}(x_n;\widehat{r}_n,\mu^r_n)}{\int_{x}p_{\mathcal{X}}(x;\bm{\alpha})\mathcal{N}(x;\widehat{r}_n,\mu^r_n)}.
\end{eqnarray}

Using the Laplace distribution, $p_{\mathcal{X}}(x_n;b)$, as in (\ref{Laplace_distribution}) we first establish in Appendix A the following result:
\begin{eqnarray}\label{posterior_numerator}
\!\!\!\!\!\!\!\!\!p_{\mathcal{X}}(x_n;b)\mathcal{N}(x_n;\widehat{r}_n,\mu^r_n)&\!\!\!\!&\nonumber\\
&& \!\!\!\!\!\!\!\!\!\!\!\!\!\!\!\!\!\!\!\!\!\!\!\!\!\!\!\!\!=~ \frac{1}{2b}\mathlarger{e^{-\alpha_n(x_n)}}\mathcal{N}\big(x_n;\gamma_n(x_n),\!~\mu^r_n\big),
\end{eqnarray}  
where the $(\widehat{r}_n,\mu^r_n,b)-$dependent functions $\alpha_n(x)$ and $\gamma_n(x)$ are given by:
\begin{eqnarray}\label{alpha_function}
\alpha_n(x)&=& \sgn(x)\frac{\widehat{r}_n}{b}~ -~ \frac{\mu^r_n}{2b^2},\\
\label{gamma_function}\gamma_n(x)&=&  \widehat{r}_n~-~\sgn(x)\frac{\mu^r_n}{b}.   
\end{eqnarray}
In (\ref{alpha_function}) and (\ref{gamma_function}), $\sgn(x)$ is the standard signum function given by:
\begin{eqnarray}\label{signum_function}
\sgn(x) &=& \begin{cases}
~1 &\textrm{if}~ x>0,\\
~0 &\textrm{if}~ x=0,\\
-1&\textrm{if}~ x<0.
\end{cases}
\end{eqnarray} 

Plugging (\ref{posterior_numerator}) back into (\ref{psoterior_prob}), it follows that:
\begin{eqnarray}\label{approximate_posterior}
p_{\mathcal{X}_n|\bm{\mathcal{Y}}}(x_n|\mathbf{y};\widehat{r}_n,\mu^r_n,\bm{\theta})&\!\!=\!\!&\frac{\mathlarger{e^{-\alpha_n(x_n)}}}{2b\psi_n}\mathcal{N}\big(x_n;\!~\gamma_n(x_n),\!~\mu^r_n\big),\nonumber\\
&\!\!\!\!&
\end{eqnarray} 
in which the normalization factor $\psi_n$ is given by:  
\begin{eqnarray}
\!\!\!\!\psi_n&=&\frac{1}{2b}\int_{-\infty}^{+\infty} \mathlarger{e^{-\alpha_n(x)}}\mathcal{N}\big(x;\!~\gamma_n(x),\!~\mu^r_n\big)\,\mathrm{d}x.
\end{eqnarray}

Using (\ref{alpha_function}) to (\ref{signum_function}), we  show after some algebraic manipulations that:
\begin{eqnarray}\label{Psi_n}
\!\!\!\!\!\!\!\!\!\psi_n&\!\!=\!\!& \frac{1}{2b}\left[e^{-\alpha_n^-}\textsf{Q}\left(\frac{\gamma_n^-}{\sqrt{\mu^r_n}}\right)\,+\,e^{-\alpha_n^+}\textsf{Q}\left(-\frac{\gamma_n^+}{\sqrt{\mu^r_n}}\right)\right]\!,
\end{eqnarray}
  wherein $\textsf{Q}(.)$ is the standard $Q$-function, i.e., the tail probability of the standard normal distribution: 
\begin{eqnarray}
\textsf{Q}(x)&\triangleq&\frac{1}{\sqrt{2\pi}}\int_{x}^{+\infty}\mathlarger{e^{-\frac{u^2}{2}}}\,\mathrm{d}u,
\end{eqnarray} 
and the $(\widehat{r}_n,\mu^r_n,b)-$dependent quantities, $\alpha_n^-$, $\alpha_n^+$, $\gamma_n^-$, and $\gamma_n^+$ are given by:
\begin{eqnarray}\label{alpha_minus}
\alpha_n^-&=& -\frac{\widehat{r}_n}{b}~ -~ \frac{\mu^r_n}{2b^2},\\
\label{alpha_plus}\alpha_n^+&=& \frac{\widehat{r}_n}{b}~ -~ \frac{\mu^r_n}{2b^2},\\
\label{gamma_minus}\gamma_n^-&=& \widehat{r}_n~+~\frac{\mu^r_n}{b},  \\
\label{gamma_plus}\gamma_n^+&=& \widehat{r}_n~-~\frac{\mu^r_n}{b}.
\end{eqnarray}

Now, the posterior mean involved in Line 15 of Algorithm 1  is given by\footnote{\changeb{Note that the dependence of $\widehat{r}_n$ and $\mu_n^r$ on the iteration index $t$ will be dropped from now on to ease notations.}}:
\begin{eqnarray}
\widehat{x}_n(t+1)&\!\!=\!\!&\int_{\mathbb{R}}x_np_{\mathcal{X}_n|\bm{\mathcal{Y}}}(x_n|\mathbf{y};\widehat{r}_n,\mu^r_n,\bm{\theta})\,\mathrm{d}x_n,
\end{eqnarray}
and owing to (\ref{approximate_posterior}), it follows that:
\begin{eqnarray}
\!\!\widehat{x}_n(t+1)&\!\!\!\!=\!\!\!\!&\frac{1}{2b\psi_n}\!\int_{\mathbb{R}}\!x_n\mathlarger{e^{-\alpha_n(x_n)}}\mathcal{N}\big(x_n;\gamma_n(x_n),\mu^r_n\big)\,\mathrm{d}x_n.\nonumber\\
\end{eqnarray}

Then, by using (\ref{alpha_function})-(\ref{gamma_function}) along with (\ref{alpha_minus})-(\ref{gamma_plus}) and resorting to some algebraic manipulations, we establish the following result:
\begin{eqnarray}\label{AppendixA_1}
\!\!\!\!\!\!\!\!\widehat{x}_n(t+1)&&\nonumber\\
&&\!\!\!\!\!\!\!\!\!\!\!\!\!\!\!\!\!\!\!\!\!\!\!\!\!\!\!\!\!\!\!\!=\frac{1}{2b\psi_n}\bigg[\mathlarger{e}^{-\alpha_n^+}\Phi_1\big(\gamma_n^+,\mu^r_n\big) \,-\,\mathlarger{e}^{-\alpha_n^-}\Phi_1\big(\!-\gamma_n^-,\mu^r_n\big) \bigg]\!,
\end{eqnarray}
in which  $\Phi_1(\gamma,\mu)$ is defined as follows:
\begin{eqnarray}\label{Phi_1}
\Phi_1\big(\gamma,\mu\big)&\triangleq& \frac{1}{\sqrt{2\pi\mu}}\int_{0}^{+\infty}t\mathlarger{e^{-\frac{(t-\gamma)^2}{2\mu}}}dt.
\end{eqnarray}

Moreover, it can be shown that $\Phi_1\big(\gamma,\mu\big)$ is analytically expressed as follows:
\begin{eqnarray}\label{AppendixA_2}
\Phi_1\big(\gamma,\mu\big) &=& \gamma\textsf{Q}\left(-\gamma/\sqrt{\mu}~\!\right)~+~\frac{\mu}{\sqrt{2\pi\mu}}\mathlarger{e^{-\frac{\gamma^2}{2\mu}}}.
\end{eqnarray}
Injecting (\ref{AppendixA_2}) in (\ref{AppendixA_1}) an rearranging the terms, it is straightforward to show that:
\begin{eqnarray}\label{AppandixA_3}
\widehat{x}_n(t+1) && \nonumber\\&&\!\!\!\!\!\!\!\!\!\!\!\!\!\!\!\!\!\!\!\!\!\!\!\!=\,\frac{1}{2b\psi_n}\Bigg[\mathlarger{e}^{-\alpha_n^-}\gamma_n^-\textsf{Q}\left(\frac{\gamma_n^-}{\sqrt{\mu^r_n}}\right)~+~\mathlarger{e}^{-\alpha_n^+}\gamma_n^+\textsf{Q}\left(\frac{-\gamma_n^+}{\sqrt{\mu^r_n}}\right)\nonumber\\
&&\!\!\!\!\!\!\!\!\!\!\!\!+ \,\frac{\mu^r_n}{\sqrt{2\pi\mu^r_n}}\left(\mathlarger{e^{-\alpha_n^+-\frac{(\gamma_n^+)^2}{2\mu^r_n}}}-\,\mathlarger{e^{-\alpha_n^--\frac{(\gamma_n^-)^2}{2\mu^r_n}}}\right)\!\Bigg]\!.
\end{eqnarray}

We also establish the following identity by combining (\ref{alpha_minus})-(\ref{gamma_plus}):
\begin{eqnarray}\label{Appendix_A_identity}  
\alpha_n^+~+~\frac{(\gamma_n^+)^2}{2\mu^r_n} &=& \alpha_n^-~+~\frac{(\gamma_n^-)^2}{2\mu^r_n}~\,=\,~\frac{\widehat{r}_n^2}{2\mu_n^r}.
\end{eqnarray}
This cancels the last two terms in (\ref{AppandixA_3})  thereby leading to the following  simple expression for the posterior mean of $\mathcal{X}_n$:
\begin{eqnarray}\label{Theorem_mean}
\widehat{x}_n(t+1) && \nonumber\\&&\!\!\!\!\!\!\!\!\!\!\!\!\!\!\!\!\!\!\!\!\!\!\!\!=\frac{1}{2b\psi_n}\left[\mathlarger{e}^{-\alpha_n^-}\gamma_n^-\textsf{Q}\left(\frac{\gamma_n^-}{\sqrt{\mu^r_n}}\right)+\mathlarger{e}^{-\alpha_n^+}\gamma_n^+\textsf{Q}\left(\frac{-\gamma_n^+}{\sqrt{\mu^r_n}}\right)\right]\!\!.\nonumber\\
\end{eqnarray} 

Now,  the posterior variance in Line 16 of Algorithm 1 is given by:
\begin{eqnarray}\label{posterior_variance}
\mu^x_n(t+1) &=&\sigma_{\mathcal{X}_n}^2\!(t+1)~-~\widehat{x}_n(t+1)^2,
\end{eqnarray}
where $\sigma_{\mathcal{X}_n}^2\!(t+1)$ is the posterior second moment of $\mathcal{X}_n$ at iteration $t+1$:
\begin{eqnarray} 
\!\!\!\!\!\!\!\sigma_{\mathcal{X}_n}^2\!(t+1)&\triangleq&\, \textsf{E}_{\mathcal{X}_n|\bm{\mathcal{Y}}}\Big\{x_n^2|\mathbf{y};\widehat{r}_n(t),\mu^r_n(t),\bm{\theta}\Big\}.
\end{eqnarray}

Using the posterior distribution in (\ref{approximate_posterior}), $\sigma_{\mathcal{X}_n}^2\!(t+1)$ is given by: 
\begin{eqnarray}
\sigma_{\mathcal{X}_n}^2(t+1)&\!\!\!\!=\!\!\!\!&   \frac{1}{2b\psi_n}\!\int_{\mathbb{R}}\!x_n^2\mathlarger{e^{-\alpha_n(x_n)}}\mathcal{N}\Big(x_n;\gamma_n(x_n),\mu^r_n\Big)\,\mathrm{d}x_n.\nonumber\\
\end{eqnarray}
whose analytical expression is also established in Appendix B as follows:
\begin{eqnarray}\label{AppendixA_9}
\!\!\!\!\!\!\!\!\sigma_{\mathcal{X}_n}^2(t+1)&\!\!=\!\!& \frac{1}{2b\psi_n}\Bigg[ \Big((\gamma_n^+)^2+\mu^r_n\Big)\mathlarger{e}^{-\alpha_n^+}\textsf{Q}\left(\!\frac{-\gamma_n^+}{\sqrt{\mu^r_n}}\right)\nonumber\\
&\!\!\!\!&~~~~+\,\Big((\gamma_n^-)^2+\mu^r_n\Big) \mathlarger{e}^{-\alpha_n^-}\textsf{Q}\left(\!\frac{\gamma_n^-}{\sqrt{\mu^r_n}}\right)\nonumber\\
&\!\!\!\!&~~~~~~~~~~~~~~~~~~~~-\,\frac{2(\mu^r_n)^2}{b\sqrt{2\pi\mu^r_n}}\mathlarger{e^{-\frac{\widehat{r}_n^2}{2\mu_n^r}}}\Bigg]\!, 
\end{eqnarray} 
Plugging (\ref{AppendixA_9}) back in (\ref{posterior_variance}) and using  (\ref{Theorem_mean}) yields the required $(t+1)th$ update for the $n$th posterior variance, $\mu^x_n(t+1)$,  in Algorithm 1. 
 
Recall also that the quantities $\alpha_n^-$, $\alpha_n^+$, $\gamma_n^-$, and $\gamma_n^+$ that are required to evaluate both $\widehat{x}_n(t+1)$ and $\mu^x_n(t+1)$ all depend on the scale parameter $b$ of the prior Laplacian distribution and the noise variance $\sigma_w^2$. Since these two parameters are also unknown  beforehand, in the next Section,  we devise an EM-based maximum-likelihood (ML) approach that learns them from the received data $\mathbf{y}$. The proposed iterative ML approach runs side by side with GAMP  wherein the soft outputs  of the latter  are used to progressively refine the EM-based estimates for $b$ and $\sigma_w^2$ and vice versa.  

\section{Learning the Scale Parameter of the Prior Laplacian Distribution and the Noise Variance}
The problem addressed in this Section is to find the ML estimate, $\widehat{\bm{\theta}}$, of $\bm{\theta}\triangleq[b,\sigma_w]$ given solely the set of recorded data:
\begin{eqnarray}\label{objective_function}
\widehat{\bm{\theta}}&=& \argmax_{\bm{\theta}}\,\ln p_{\bm{\mathcal{Y}}}(\mathbf{y};\bm{\theta}),
\end{eqnarray} 
where $p_{\bm{\mathcal{Y}}}(\mathbf{y};\bm{\theta})$ is the pdf of $\bm{\mathcal{Y}}$ parameterized by  $\bm{\theta}$ given by:
\begin{eqnarray}
p_{\bm{\mathcal{Y}}}(\mathbf{y};\bm{\theta})&=&\int_{\mathbf{x}}p_{\bm{\mathcal{Y}}|\bm{\mathcal{X}}}(\mathbf{y}|\mathbf{x};\sigma_w)p(\mathbf{x};b)\,\mathrm{d}\mathbf{x}.
\end{eqnarray}

The objective function in (\ref{objective_function}) is  a nonlinear transformation of the parameters and its analytical maximization is mathematically intractable. Yet, iterative solutions can be envisaged to solve the underlying optimization problem numerically. In this paper, we resort to the EM concept \cite{EM} which is a widely used tool in ML vector parameter estimation practices. More interestingly,  GAMP  returns the adequate posterior probabilities that are  required by the EM algorithm in order to update its estimates. The successful formulation of the EM algorithm amounts to the appropriate identification of the so-called \textit{incomplete} and \textit{complete} data sets that are adequate to the estimation problem at hand. In our case, they are taken to be $\mathbf{y}$ and $\mathbf{v} \triangleq [\mathbf{y}, \mathbf{x}]^T$, respectively. Then, instead of maximizing the actual log-likelihood function (LLF), $\ln p_{\bm{\mathcal{Y}}}(\mathbf{y};\bm{\theta})$, the EM algorithm maximizes $\ln p_{\bm{\mathcal{V}}}(\mathbf{v};\bm{\theta})$.  Since the set of complete data, $\mathbf{v}$, is not entirely available, the EM algorithm replaces  $\ln p_{\bm{\mathcal{V}}}(\mathbf{v};\bm{\theta})$ by its conditional expectation: 
\begin{eqnarray}\label{expected_LLF_EM} 
\mathbb{E}_{\bm{\mathcal{V}}|\bm{\mathcal{Y}}}\Big\{\ln p_{\bm{\mathcal{V}}}(\mathbf{v};\bm{\theta})|\mathbf{y}\Big\}&\!\!\!\!=\!\!\!\!&\int_{\mathbf{v}}\ln p_{\bm{\mathcal{V}}}(\mathbf{v};\bm{\theta}) p_{\bm{\mathcal{V}}|\bm{\mathcal{Y}}}(\mathbf{v}|\mathbf{y};\bm{\theta})\,\mathrm{d}\mathbf{v}.\nonumber\\
&\!\!\!\!\!\!\!\!&
\end{eqnarray}
However, since one needs to know $\bm{\theta}$ in order to determine $p_{\bm{\mathcal{V}}|\bm{\mathcal{Y}}}(\mathbf{v}|\mathbf{y};\bm{\theta})$ and hence the expected LLF in (\ref{expected_LLF_EM}), the EM algorithm proceeds in the following iterative way.  We start with an initial guess, $\widehat{\bm{\theta}}^{(0)}$, about the unknown parameter vector $\bm{\theta}$, and we let $\widehat{\bm{\theta}}_k$ be the $k^{th}$ guess of its MLE. 
 Then, as the name suggests, the expectation-maximization algorithm alternates between the following two main steps:

$\bullet$ \textbf{Expectation step (\textsc{E-STEP})}: Find the average log-likelihood of the complete data:
\begin{eqnarray}\label{E-step}
\!\!\!\!\!\!Q(\bm{\theta},\widehat{\bm{\theta}}_k)&\!\!=\!\!&\int_{\mathbf{v}}\ln p_{\bm{\mathcal{V}}}(\mathbf{v};\bm{\theta}) p_{\bm{\mathcal{V}}|\bm{\mathcal{Y}}}\big(\mathbf{v}|\mathbf{y};\widehat{\bm{\theta}}_k\big)\,\mathrm{d}\mathbf{v}.
\end{eqnarray}

$\bullet$ \textbf{Maximization step (\textsc{M-STEP})}: Maximize the average log-likelihood of the complete data:
\begin{eqnarray}\label{M-step}
\widehat{\bm{\theta}}_{k+1}&=& \argmax_{\bm{\theta}}\,Q(\bm{\theta},\widehat{\bm{\theta}}_k).
\end{eqnarray}
 The algorithm stops once the user-specified convergence criterion  $|\widehat{\bm{\theta}}_{k+1}-\widehat{\bm{\theta}}_{k}|\leq \delta$ is met or a predefined maximum number of iterations is attained; whichever occurs first. 
 
 Now, recalling that $\mathbf{v} \triangleq [\mathbf{y}, \mathbf{x}]^T$ and using the Bayes' rule, we write:
 \begin{equation}\label{Bayes_EM}
 ~~p_{\bm{\mathcal{V}}}(\mathbf{v};\bm{\theta})~=~ p_{\bm{\mathcal{X}},\bm{\mathcal{Y}}}(\mathbf{y},\mathbf{x};\bm{\theta})~=~p_{\bm{\mathcal{Y}}|\bm{\mathcal{X}}}(\mathbf{y}|\mathbf{x};\sigma_w^2)p_{\bm{\mathcal{X}}}(\mathbf{x};b).
 \end{equation}  
 Plugging (\ref{Bayes_EM}) back into (\ref{E-step}) and using:
 \begin{eqnarray}
 p_{\bm{\mathcal{V}}|\bm{\mathcal{Y}}}\big(\mathbf{v}|\mathbf{y};\widehat{\bm{\theta}}_k\big) &=& p_{\bm{\mathcal{X}}|\bm{\mathcal{Y}}}\big(\mathbf{x}|\mathbf{y};\widehat{\bm{\theta}}_k\big),
 \end{eqnarray} 
it can be shown  that:
\begin{eqnarray}\label{E-step1}
\!\!\!\!\!\!\!\!\!\!Q(\bm{\theta},\widehat{\bm{\theta}}_k)&\!\!=\!\!&\int_{\mathbf{x}}\ln p_{\bm{\mathcal{Y}}|\bm{\mathcal{X}}}(\mathbf{y}|\mathbf{x};\sigma_w^2) p_{\bm{\mathcal{X}}|\bm{\mathcal{Y}}}\big(\mathbf{x}|\mathbf{y};\widehat{\bm{\theta}}_k\big)\,\mathrm{d}\mathbf{x}\nonumber\\
&&~~~~+\,\int_{\mathbf{x}}\ln p_{\bm{\mathcal{X}}}(\mathbf{x};b) p_{\bm{\mathcal{X}}|\bm{\mathcal{Y}}}\big(\mathbf{x}|\mathbf{y};\widehat{\bm{\theta}}_k\big)\,\mathrm{d}\mathbf{x}.
\end{eqnarray}

As seen from (\ref{E-step1}), $Q(\bm{\theta},\widehat{\bm{\theta}}_k)$ decouples in terms of the unknown parameters $\sigma_w$ and $b$. Therefore, the $k^{tk}$ EM update $\widehat{\bm{\theta}}_{k+1} = [\widehat{\sigma}^2_{w,k+1},\widehat{b}_{k+1}]$  of $\bm{\theta} = [\sigma_w^2~b]$ given  in (\ref{M-step}) is obtained as follows:
\begin{eqnarray}
\!\!\!\!\!\!\!\!\!\label{argmax_b}\widehat{b}_{k+1}&\!\!\!=\!\!&\argmax_{b>0}\mathbb{E}_{\bm{\mathcal{X}}|\bm{\mathcal{Y}}}\Big\{\ln p_{\bm{\mathcal{X}}}(\mathbf{x};b)\big|\mathbf{y}, \widehat{\bm{\theta}}_{k}\Big\}.\\
\label{argmax_sigma}
\!\!\!\!\!\!\!\!\!\widehat{\sigma}^2_{w,k+1} &\!\!\!=\!\!& \argmax_{\changeb{\sigma^2_w>0}}\mathbb{E}_{\bm{\mathcal{X}}|\bm{\mathcal{Y}}}\Big\{\ln p_{\bm{\mathcal{Y}}|\bm{\mathcal{X}}}(\mathbf{y}|\mathbf{x};\sigma^2_w)\big|\mathbf{y}, \widehat{\bm{\theta}}_{k}\Big\},
\end{eqnarray}

We further assume the elements of $\mathbf{x}$ to be i.i.d.:
\begin{eqnarray}
p_{\bm{\mathcal{X}}}(\mathbf{x};b)&=&\prod_{n=1}^{N}p_{\mathcal{X}}(x_n;b).
\end{eqnarray}
Hence, we rewrite (\ref{argmax_b}) as follows:
\begin{eqnarray}\label{argmax_b_0}
\!\!\!\!\!\!\!\!\!\widehat{b}_{k+1}
&\!\!=\!\!& \argmax_{b>0}\sum_{n=1}^{N}\mathbb{E}_{\bm{\mathcal{X}}|\bm{\mathcal{Y}}}\left\{\ln p_{\mathcal{X}}(x_n;b)\big|\mathbf{y};\widehat{\bm{\theta}}_k\right\},
\end{eqnarray}

Recall from (\ref{E-step1}) that the expectation in (\ref{argmax_b_0}) is taken with respect to the posterior density $p_{\bm{\mathcal{X}}|\bm{\mathcal{Y}}}\big(\mathbf{x}|\mathbf{y};\widehat{\bm{\theta}}_k\big)$ which we further approximate by:
 \begin{eqnarray}\label{posterior_approx}
 p_{\bm{\mathcal{X}}|\bm{\mathcal{Y}}}\big(\mathbf{x}|\mathbf{y};\widehat{\bm{\theta}}_k\big)&=&\prod_{n=1}^{N}p_{\mathcal{X}_n|\bm{\mathcal{Y}}}\big(x_n|\mathbf{y};\widehat{\bm{\theta}}_k\big).
 \end{eqnarray}
 Using (\ref{posterior_approx}) in (\ref{argmax_b_0}), it follows that: 
\begin{eqnarray}\label{argmax_b_2}
\!\!\!\!\!\!\!\!\!\widehat{b}_{k+1}&\!\!=\!\!& \argmax_{b>0}\underbrace{\sum_{n=1}^{N}\mathbb{E}_{\mathcal{X}_n|\bm{\mathcal{Y}}}\left\{\ln p_{\mathcal{X}}(x_n;b)\big|\mathbf{y};\widehat{\bm{\theta}}_k\right\}}_{\substack{\\\mathlarger{\triangleq}\\\mathlarger{g(b|\mathbf{y},\bm{\widehat{\theta}}_k)}}}.
\end{eqnarray}
Clearly, $\widehat{b}_{k+1}$ is the value of $b$ that zeros the first derivative  of  $g(b|\mathbf{y},\widehat{\bm{\theta}}_k)$ with respect to $b$:
\begin{eqnarray} \label{g_first_derivative}
\frac{\mathrm{d}}{\mathrm{d}b}g(b|\mathbf{y},\widehat{\bm{\theta}}_k)&\!\!\!=\!\!\!&\sum_{n=1}^{N}\mathbb{E}_{\mathcal{X}_n|\bm{\mathcal{Y}}}\left\{\frac{1}{p_{\mathcal{X}}(x_n;b)}\frac{\mathrm{d}}{\mathrm{d}b} p_{\mathcal{X}}(x_n;b)\Big|\mathbf{y};\widehat{\bm{\theta}}_k\right\}.\nonumber\\
&&
 \end{eqnarray}

It is also easy to show the following result for the Laplacian distribution given in (\ref{Laplace_distribution}):
\begin{eqnarray}
\frac{\mathrm{d}}{\mathrm{d}b} p_{\mathcal{X}}(x_n;b)&=& \frac{1}{b}\left(\frac{|x_n|}{b}-1\right)p_{\mathcal{X}}(x_n;b),
\end{eqnarray}
which is injected back into (\ref{g_first_derivative}) to obtain:
\begin{eqnarray} \label{g_first_derivative_1}
\!\!\!\!\!\!\!\!\frac{\mathrm{d}}{\mathrm{d}b}g(b|\mathbf{y},\widehat{\bm{\theta}}_k)&\!\!=\!\!&\frac{1}{b}\sum_{n=1}^{N}\mathbb{E}_{\mathcal{X}_n|\bm{\mathcal{Y}}}\left\{\frac{|x_n|}{b}-1\Big|\mathbf{y};\widehat{\bm{\theta}}_k\right\}.
 \end{eqnarray}
Setting (\ref{g_first_derivative_1}) equal to zero and solving for $b$ yields the following EM update for $b$:
\begin{eqnarray} \label{EM_update_b} 
\widehat{b}_{k+1}&=& \frac{1}{N}\sum_{n=1}^{N}\mathbb{E}_{\mathcal{X}_n|\bm{\mathcal{Y}}}\left\{\,|x_n|\,\Big|\,\mathbf{y};\widehat{\bm{\theta}}_k\right\} 
\end{eqnarray} 

The posterior first-order moment in (\ref{EM_update_b}) for each $x_n$  is computed using  the associated    posterior distribution $p_{\mathcal{X}_n|\bm{\mathcal{Y}}}\big(x_n|\mathbf{y};\widehat{\bm{\theta}}_k\big)$. The latter is readily obtained from the auxiliary outputs of GAMP according to (\ref{approximate_posterior}) wherein the true parameter vector $\bm{\theta}$ is now replaced by its previous EM update $\widehat{\bm{\theta}}_{k}$. More specifically,  we have:
\begin{eqnarray}\label{approximate_posterior_EM}
\!\!\!\!\!\!\!\!p_{\mathcal{X}_n|\bm{\mathcal{Y}}}(x_n|\mathbf{y};\widehat{r}_n,\mu^r_n,\widehat{\bm{\theta}}_k)&&\nonumber\\
&&\!\!\!\!\!\!\!\!\!\!\!\!\!\!\!\!\!\!\!\!\!\!\!\!\!\!\!\!\!\!\!=\,\frac{\mathlarger{e^{-\alpha_{n,k}(x_n)}}}{2\widehat{b}_k\psi_{n,k}}\mathcal{N}\big(x_n;\!~\gamma_{n,k}(x_n),\!~\mu^r_n\big), 
\end{eqnarray} 
where the functions $\alpha_{n,k}(x)$ and $\gamma_{n,k}(x)$ are obtained from (\ref{alpha_function}) and (\ref{gamma_function}), respectively, with $b$ being replaced by $\widehat{b}_k$. Likewise, $\psi_{n,k}$ has the same expression as in (\ref{Psi_n}); the only difference being that $b$ is replaced by $\widehat{b}_k$ in all the $b-$dependent quantities $\alpha_n^-$, $\alpha_n^+$, $\gamma_n^-$, and $\gamma_n^+$. For convenience,  we will also  from now on denote these quantities as  $\alpha_{n,k}^-$, $\alpha_{n,k}^+$, $\gamma_{n,k}^-$, and $\gamma_{n,k}^+$, respectively. 
Now, using the EM posterior distribution in (\ref{approximate_posterior_EM}), it can be further shown that: 
\begin{eqnarray} \label{posterior_moment}
\!\!\!\!\!\!\!\!\!\!\mathbb{E}_{\mathcal{X}_n|\bm{\mathcal{Y}}}\left\{\,|x_n|\,\big|\,\mathbf{y};\widehat{\bm{\theta}}_k\right\} &\!\!\!\!\!\!&\nonumber\\
&&\!\!\!\!\!\!\!\!\!\!\!\!\!\!\!\!\!\!\!\!\!\!\!\!\!\!\!\!\!\!\!\!\!\!\!\!\!\!\!\!\!=\frac{1}{2\widehat{b}_k\psi_{n.k}}\bigg[\mathlarger{e}^{-\alpha_{n,k}^+}\Phi_1\big(\gamma_{n,k}^+,\mu^r_n\big) \nonumber\\
&&+~\mathlarger{e}^{-\alpha_{n,k}^-}\Phi_1\big(\!-\gamma_{n,k}^-,\mu^r_n\big) \bigg]\!.
\end{eqnarray}

 \begin{figure*}[t]    
 \vskip -1.5cm
\begin{centering}
\includegraphics[scale=0.4]{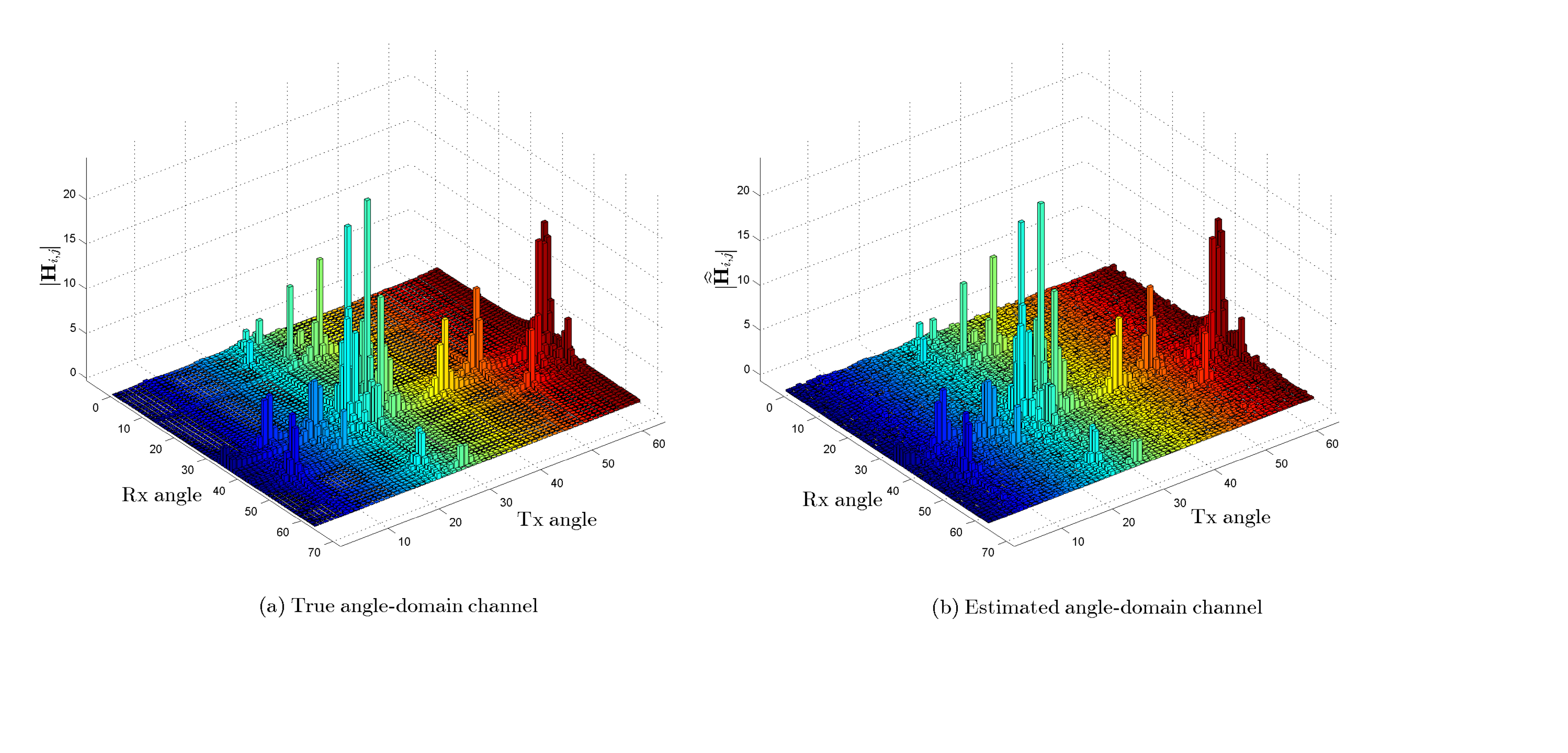}
 \vskip -1.5cm
\caption{(a) True and estimated angular-domain channel for $M_t=64$, $M_r=64$, $K=128$, and $\textrm{SNR}=10$ dB. Channel had 4 clusters each with angular spread  3.5 degrees and consisting of 10 subpaths: (a) true channel and (b) GAMP-Laplace estimated channel.} 
\label{true_VS_estimated_channel}
\end{centering}
\end{figure*}

Then, after \changeb{using} (\ref{AppendixA_2}) and  (\ref{Appendix_A_identity}) in (\ref{posterior_moment}), and recalling (\ref{EM_update_b}),  we establish the EM update for the scale parameter, $b$, as follows: 
  \begin{eqnarray}
  \widehat{b}_{k+1} &\changeb{=}&\!\!\frac{1}{N}\sum_{n=1}^N\frac{1}{\xi_{n,k}}\frac{2\mu_n^r}{\sqrt{2\pi\mu_n^r}}\mathlarger{e^{-\frac{\widehat{r}_n^2}{2\mu_n^r}}}+\frac{1}{N}\sum_{n=1}^N\frac{\xi_{n,k}'}{\xi_{n,k}},\nonumber\\
  \end{eqnarray}
in which $\xi_{n,k}$ and $\xi_{n,k}'$ are explicitly given by:
\begin{eqnarray}
\xi_{n,k}&\!\!\!=\!\!\!&~\mathlarger{e}^{-\alpha_{n,k}^+}\textsf{Q}\left(\!\frac{-\gamma_{n,k}^+}{\sqrt{\mu_n^r}}\!\right)~+~\,\mathlarger{e}^{-\alpha_{n,k}^-}\textsf{Q}\left(\frac{\gamma_{n,k}^-}{\sqrt{\mu_n^r}}\right)\!,\nonumber\\
\xi_{n,k}'&\!\!\!=\!\!\!&\gamma_{n,k}^+\mathlarger{e}^{-\alpha_{n,k}^+}\textsf{Q}\left(\!\frac{-\gamma_{n,k}^+}{\sqrt{\mu_n^r}}\!\right)
\,-~\gamma_{n,k}^-\mathlarger{e}^{-\alpha_{n,k}^-}\textsf{Q}\left(\frac{\gamma_{n,k}^-}{\sqrt{\mu_n^r}}\right)\!.\nonumber\
\end{eqnarray}

In addition, since the original complex noise components in (\ref{system_model_vec_generic_0}) are assumed to be independent and modeled by circularly-symmetric Gaussian RVs with variance $2\sigma_w^2$, then the real-valued noise components, $w_m$,  in (\ref{system_model_vec_generic}) are also independent with  $w_m\sim\mathcal{N}(w_m;0,\sigma_w^2)$. Therefore, after dropping the constant term that does not depend on $\sigma_w^2$, we have:
\begin{eqnarray}\label{noise_variance_conditional}
\ln p_{\bm{\mathcal{Y}}|\bm{\mathcal{X}}}(\mathbf{y}|\mathbf{x};\sigma^2_w)&\!\!\!\!=\!\!\!\!& -\frac{M}{2}\ln(\sigma_w^2)-\frac{1}{2\sigma_w^2}\sum_{m=1}^M(y_m-z_m)^2,\nonumber\\
\end{eqnarray}
where $z_m$ is the $m$th component of $\mathbf{z}=\mathbf{Ax}$.
Finding the partial derivative of (\ref{noise_variance_conditional}) with respect to $\sigma_w^2$, plugging it back in (\ref{argmax_sigma}), then taking the expectation with respect to the posterior in (\ref{posterior_approx}), setting the result to zero, and solving for $\sigma_w^2$, yields the following EM update for the noise variance:  
\begin{eqnarray}\label{sigma_k}
\!\!\!\!\!\!\!\!\!\widehat{\sigma}^2_{w,k+1} &\!\!=\!\!&\frac{1}{M}\sum_{m=1}^M\Big(|y_m - \widehat{z}_m|^2~+~\mu_m^z\Big),
\end{eqnarray}
which is equivalent to the result obtained earlier in \cite{Schniter_EM_GM} under the GM model and complex data. 
Note also that a more general prior such as the elastic net prior can be used to model the beam-domain coefficients of massive MIMO mmWave  channels. The corresponding GAMP updates have been recently  established in \cite{ziniel2014message} within the framework of Bayesian logistic regression problems.

\section{Simulation Results}\label{section_5}
In this Section, we assess the performance of the proposed massive MIMO mmWave channel estimator using exhaustive Monte-Carlo simulations and the normalized mean-square error (NMSE) as a performance measure:
\begin{eqnarray}  
\textrm{NMSE} &=&\mathbb{E}\left\{\frac{\|\widehat{\mathbf{x}}-\mathbf{x}\|^2}{\|\mathbf{x}\|^2}\right\}.
\end{eqnarray} 
  
For each  $p$th cluster in (\ref{channel_model}), the gain of the corresponding $q$th sub-path  is drawn  independently from a  zero-mean complex Gaussian distribution of variance $\sigma_{p,q}^2$. The latter obey the constraint  $\sum_{q=1}^{Q_p}\sigma_{p,q}^2=\sigma_{p}^2$ which is the power fraction pertaining to the $p$th path cluster. We further normalize the total channel power by enforcing $\sum_{p=1}^{P}\sigma_{p}^2=1$ as well as the transmit power      
by enforcing $\mathbb{E}\{\|\mathbf{b}(k)\|^2\}=1$.
 Under these normalizations, it can be verified from (\ref{system_model_combining11}) that the SNR is simply given:
\begin{eqnarray}
\textrm{SNR}~\triangleq~\frac{\mathbb{E}\big\{\|\widetilde{\mathbf{H}}\mathbf{b}\|^2\big\}}{\mathbb{E}\big\{\|\widetilde{\mathbf{w}}\|^2\big\}}~=~\frac{1}{2\sigma_w^2}.
\end{eqnarray}

The AoDs (resp. AoAs) pertaining to each $p$th cluster are also drawn independently around the associated mean angle of departure (resp. arrival) with angular spread equal to 3.5 degrees. The mean angles of departure/arrival of the different path clusters are generated independently and uniformly at random in $[0,\pi]$.

Later on in this section, we also assess the performance of the proposed algorithm in terms of the achievable rate.  
As a non-Bayesian baseline, we consider the ML estimator (when $\mathbf{A}$ is full column rank) which boils down to the conventional LS estimator due to the linearity of the model in (\ref{system_model_vec_generic}) and the Gaussianity of the noise:
\begin{eqnarray}
\widehat{\mathbf{x}}_{\textrm{ML}}&=&\big(\mathbf{A}^{\textsf{T}}\mathbf{A}\big)^{-1}\mathbf{A}^{\textsf{T}}\mathbf{y},
\end{eqnarray}

From the Bayesian family, we also consider
 the GAMP-based estimator under the GM model recently investigated in \cite{AMP_mmWave} (and we refer to it simply as GAMP-GM), as well as, the well-known linear minimum mean-square error (LMMSE) estimator. We also  consider   uniform linear arrays consisting of $M_t=64$ and $M_r=64$ antenna elements both at  the transmitter and receiver sides. As a representative example, the multipath channel consists of $P=4$ clusters each of which containing $Q_p = 10$ sub-paths with the same angular spread of 3.5 degrees.  The results reported in this section are obtained using 500 Monte-Carlo trials. We also allow a maximum number of iterations $T_{\textrm{max}} = 50$ for GAMP  if it does not converge given the precision tolerance $\epsilon =10^{-6}$ (cf. algorithm 1). 
 
In all simulations, the EM algorithm was initialized as in \cite{Schniter_EM_GM} for GAMP-GM. More specifically, the noise variance was initialized as follows:
\begin{eqnarray}\label{noise_variance_init}
\widehat{\sigma}^2_{w,0}&=& \frac{\|\mathbf{y}\|_2^2}{(\textsf{SNR}^0+1)M},
\end{eqnarray}
where $M=2M_rK$ and $\textsf{SNR}^0 = 100$. Moreover, the parameters of the GM prior were initialized as follows for $l=1,\ldots,L$:
\begin{eqnarray}
\omega_l=\frac{1}{L},~~ \nu_l=\frac{l}{\sqrt{L}}\frac{\left(\|\mathbf{y}\|_2^2-M\widehat{\sigma}^2_{w,0}\right)}{\|\mathbf{A}\|^2_F},~~\textrm{and}~~\eta_l=0,
\end{eqnarray}
in which $\|.\|_F$ stands for the Frobenius norm.
For EM-GAMP-Laplace, the noise variance was initialized as in (\ref{noise_variance_init}) above and the scale parameter, $b$, of the Laplacian prior was initialized to $\widehat{b}_0=1$.

Note also that the overall stopping conditions of EM-GAMP are those corresponding to the outer iteration loop (i.e., GAMP iteration loop) shown in Line 18 of algorithm 1. The EM algorithm has its own stopping conditions under each $t^{th}$ GAMP iteration. When the stopping conditions for the EM algorithm are met, the resulting EM updates for the noise variance and the  parameters of the prior distribution are used by GAMP during its $(t+1)$th iteration which is incremented in Line 17 of Algorithm 1. 
 
To start, we illustrate  the behavior of the proposed GAMP-Laplace estimator in Fig. \ref{true_VS_estimated_channel}. We  plot  a specific realization for the true angular-domain channel's magnitude and its estimate as returned by GAMP-Laplace at $\textrm{SNR}=10$ dB. As seen from Fig. \ref{true_VS_estimated_channel}(a), the true channel is indeed approximately sparse in the angular domain, i.e.,  it exhibits a very small number of  dominant coefficients. The effect of  spectral leakage is also clearly observed as the magnitudes of the remaining coefficients become smaller as we move away from the location of the subpath clusters.  As seen from Fig. \ref{true_VS_estimated_channel}(b), GAMP-Laplace is able to  estimate the mmWave channel with high accuracy at a moderate SNR threshold.    

\begin{figure}[t]
\hskip -1cm
\begin{centering} 
\includegraphics[scale=0.56]{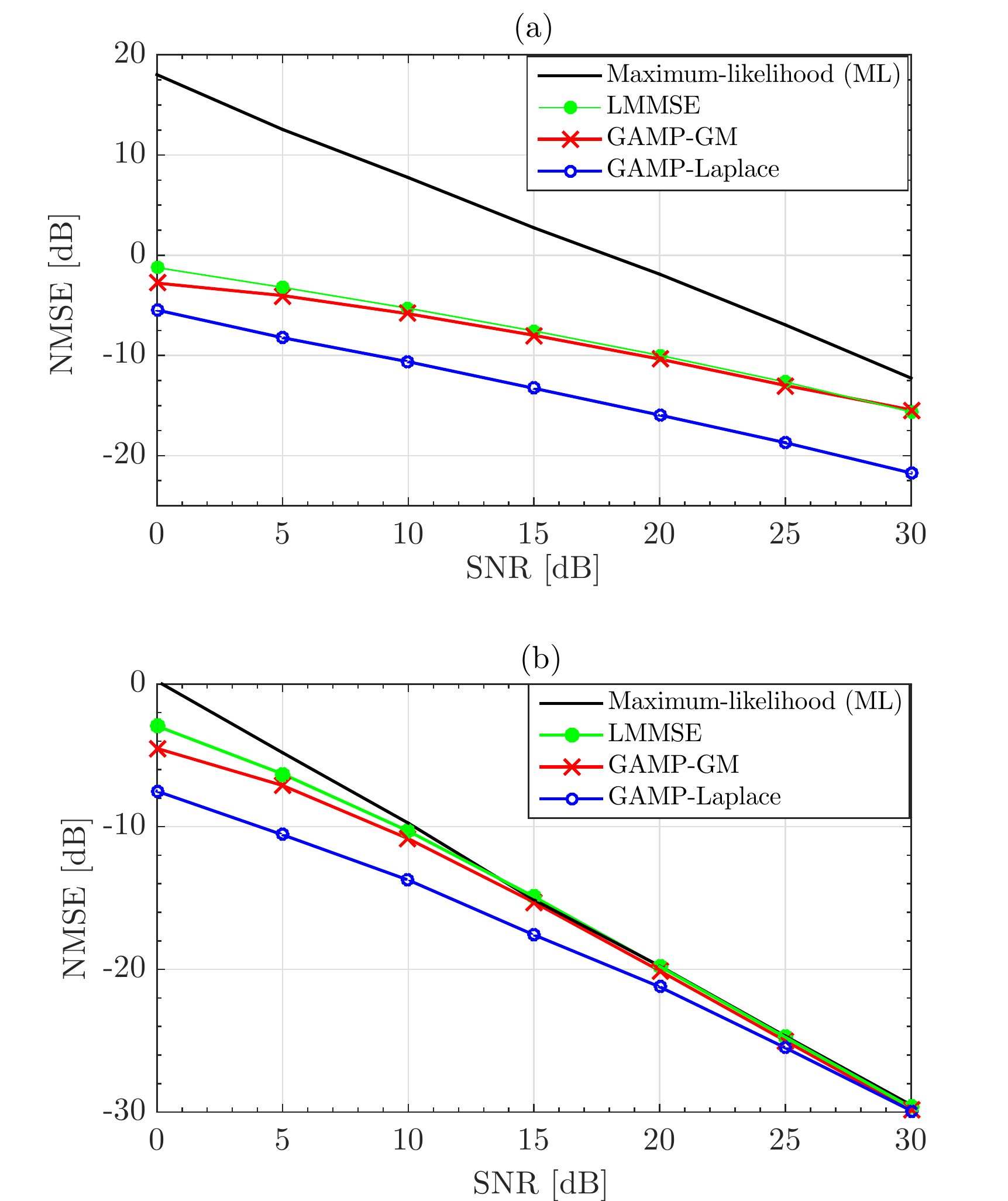} 
\caption{Estimation NMSE for GAMP-Laplace, GAMP-GM, LMMSE, and ML versus SNR: (a) $K=64$ and (b) $K=128$.}
\label{NMSE_clusters}   
 \end{centering} 
\end{figure}

Fig. \ref{NMSE_clusters} depicts the NMSE performance of the three considered estimators as function of the SNR for two different values of the observation window size, namely $K=64$ and $K=128$. First, as expected it is seen from Fig. \ref{NMSE_clusters}(a) that the performance of the ML estimator is very poor  since $K=64$  corresponds to the case where the number of unknowns $M_tM_r$ is equal to the number of observations $KM_r$. In fact, it is widely known that ML estimation, under linear mixing, requires more observations than unknowns. This is confirmed by Fig \ref{NMSE_clusters}(b) in which we double the number of observations. In the latter case, the advantage of the two Bayesian approaches over ML estimation in low-to-moderate SNRs is due to their ability to exploit the prior information about the unknown channel instead of simply assuming it to be  unknown but deterministic as is the case with ML estimation. It is also seen that in both cases GAMP-Laplace offers remarkable performance gains over GAMP-GM. For instance, at target $\textrm{NMSE}=-10$ dB, the gains in terms of SNR are as high as 10 dB and 5 dB for $K=32$ and $K=64$, respectively. As we shall see later, this translates to large gains in terms of achievable rate.

\begin{figure}[t]
\begin{centering}
\resizebox{\columnwidth}{!}{
 \includegraphics{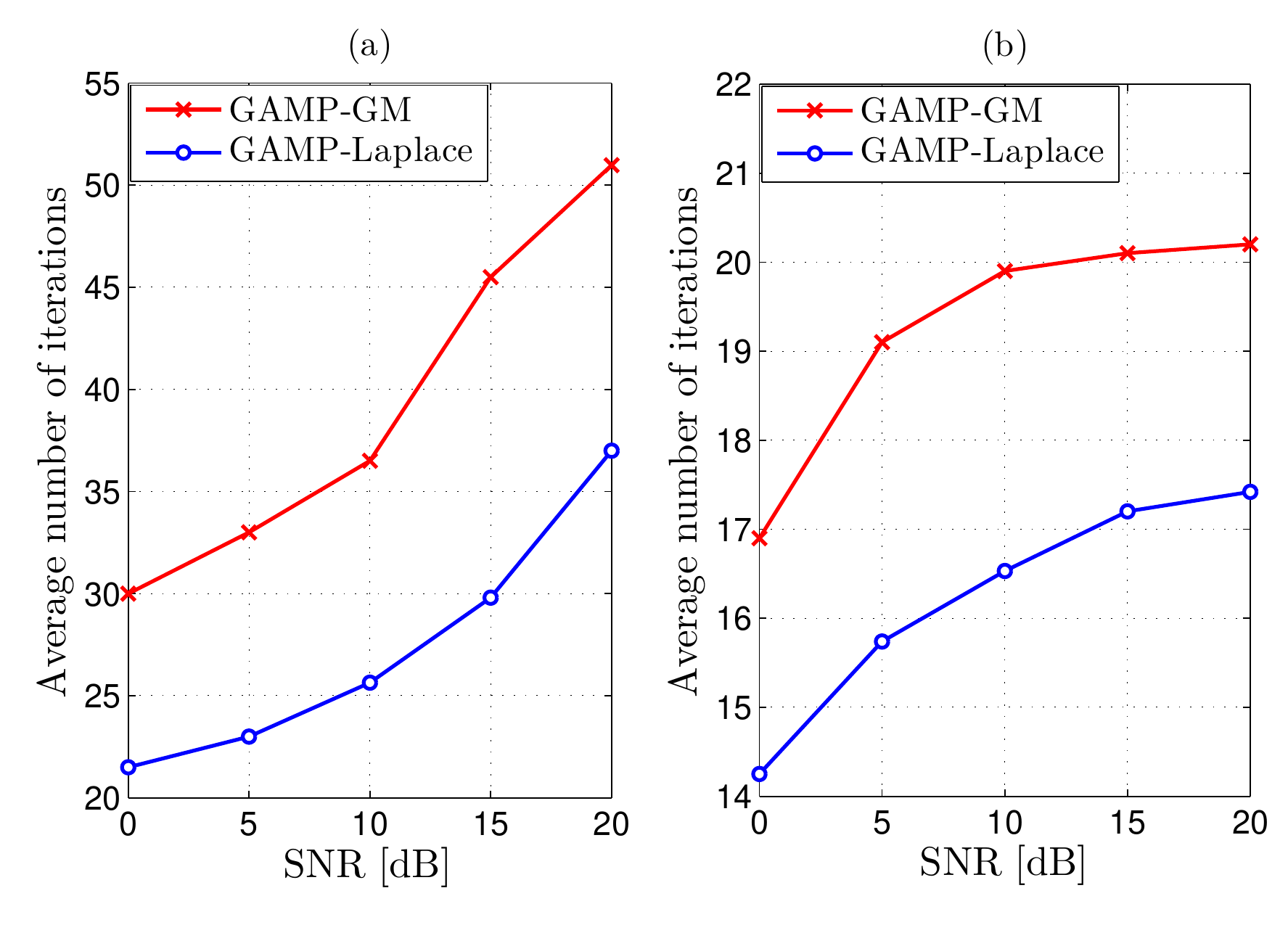}} 
  \vskip -0.25cm
\caption{Average number of iterations until convergence for GAMP-GM and GAML-Laplace: (a) $K=64$ and (b) $K=128$.}
\label{convergence_rate} 
 \end{centering}  
\end{figure}

We also plot in Fig. \ref{convergence_rate}, the  average number of iterations required by GAMP (until convergence) under both  GM and Laplace priors.  There, it is seen that  GAMP-Laplace converges much faster than  GAMP-GM due to more accurate modeling of the prior distribution of the angular-domain channel coefficients. For instance, at $\textrm{SNR}=10$ dB,  GAMP-Laplace converges in almost 25 iterations when $K=64$ as opposed to 35  iterations for GAMP-GM  thereby leading to tremendous computational savings in practice. Recall here that GAMP performs four matrix/vector multiplications at each iteration (cf. Algorithm 1 which is actually a scalarized version of GAMP). Hence, GAMP-Laplace saves on average 40 matrix/vector multiplications over GAMP-GM.  This is to be added to the computational savings stemming from  the fact that GAMP-Laplace needs to learn only one parameter (namely the scale parameter $b$), under each iteration, as opposed to $3L$ different parameters for GAMP-GM where $L$ is the order of the underlying Gaussian mixture. 

Next, we  focus on the  data decoding phase and investigate the impact of channel estimation accuracy  on the mutual information or equivalently the achievable rate. Our approach is inspired by that in \cite{AMP_mmWave,sparse_way}. We assume that the channel estimate acquired at the receiver is provided to the transmitter via feedback links, or  that the mmWave system is operating in time-division-duplex (TDD) mode with reciprocal channel which is estimated directly at the transmitter via the reverse link and used for beamforming/multiplexing purposes on the forward link. More specifically, if the estimated channel matrix, $\widehat{\mathbf{H}}$, is made available to the transmitter, the latter performs its singular-value decomposition (SVD):
\begin{eqnarray}
\widehat{\mathbf{H}}&=&\widehat{\mathbf{U}}\bm{\Sigma}\widehat{\mathbf{V}}^{\textsf{H}}.
\end{eqnarray}

\begin{figure}[t] 
\begin{centering}
\resizebox{\columnwidth}{!}{
 \includegraphics{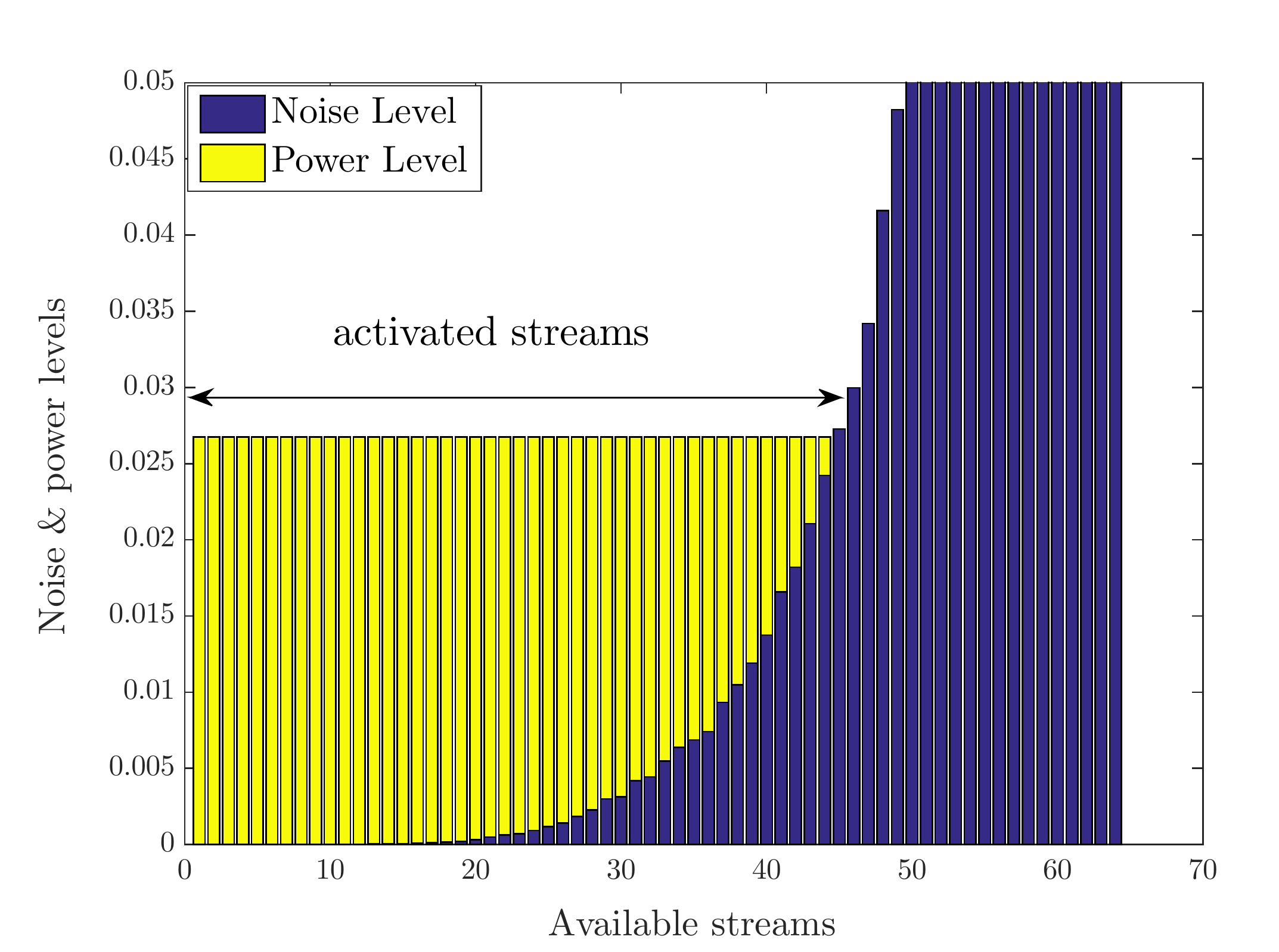}}  
  \vskip 0.25cm 
\caption{Output of the water-filling algorithm applied using GAMP-Laplace channel estimates at SNR = 10 dB.}
\label{waterfilling_figure}  
 \end{centering} 
\end{figure}

Then, $\widehat{\mathbf{V}}$ is used to produce the precoded signal: 
\begin{eqnarray}
\mathbf{s}&=& \widehat{\mathbf{V}}\,\textrm{diag}\left(\sqrt{P_1},\sqrt{P_2},\ldots,\sqrt{P_{M^*}}\right)\mathbf{a},
\end{eqnarray} 
in which $\mathbf{a}$ is the information-bearing symbol whose components contain independently coded data streams and $M^*=\textrm{min}(M_t,M_r)$.  Moreover, $\{P_m\}_{m=1}^{M^*}$ are the set of optimal powers allocated  across the various data streams as provided by the well-known water-filling algorithm. Actually, due to  the inherent  cluster-based model in (\ref{channel_model}), the underlying  mmWave channel matrix is rank deficient and hence the number (say $S$) of independent data streams it can support is much smaller than $M^*$. This is corroborated by the results reported in Fig. \ref{waterfilling_figure} which depicts the output of the water-filling algorithm when applied to the channel estimate $\widehat{\mathbf{H}}$ that is provided by  GAMP-Laplace  at $\textrm{SNR}=10$ dB. There, it is seen that nearly 40 out of the 64 available data streams are activated by the waterfilling algorithm. This is in line with the fact that the true channel matrix $\mathbf{H}$ is of rank 40 since it embodies 4 clusters each of which consisting of 10 subpaths. The fact that the activated data streams in Fig.  \ref{waterfilling_figure}  slightly exceeds the rank of  $\mathbf{H}$ is due to the channel estimation error induced by GAMP-Laplace.

 \begin{figure}[t]    
\begin{centering}
\includegraphics[scale=0.35]{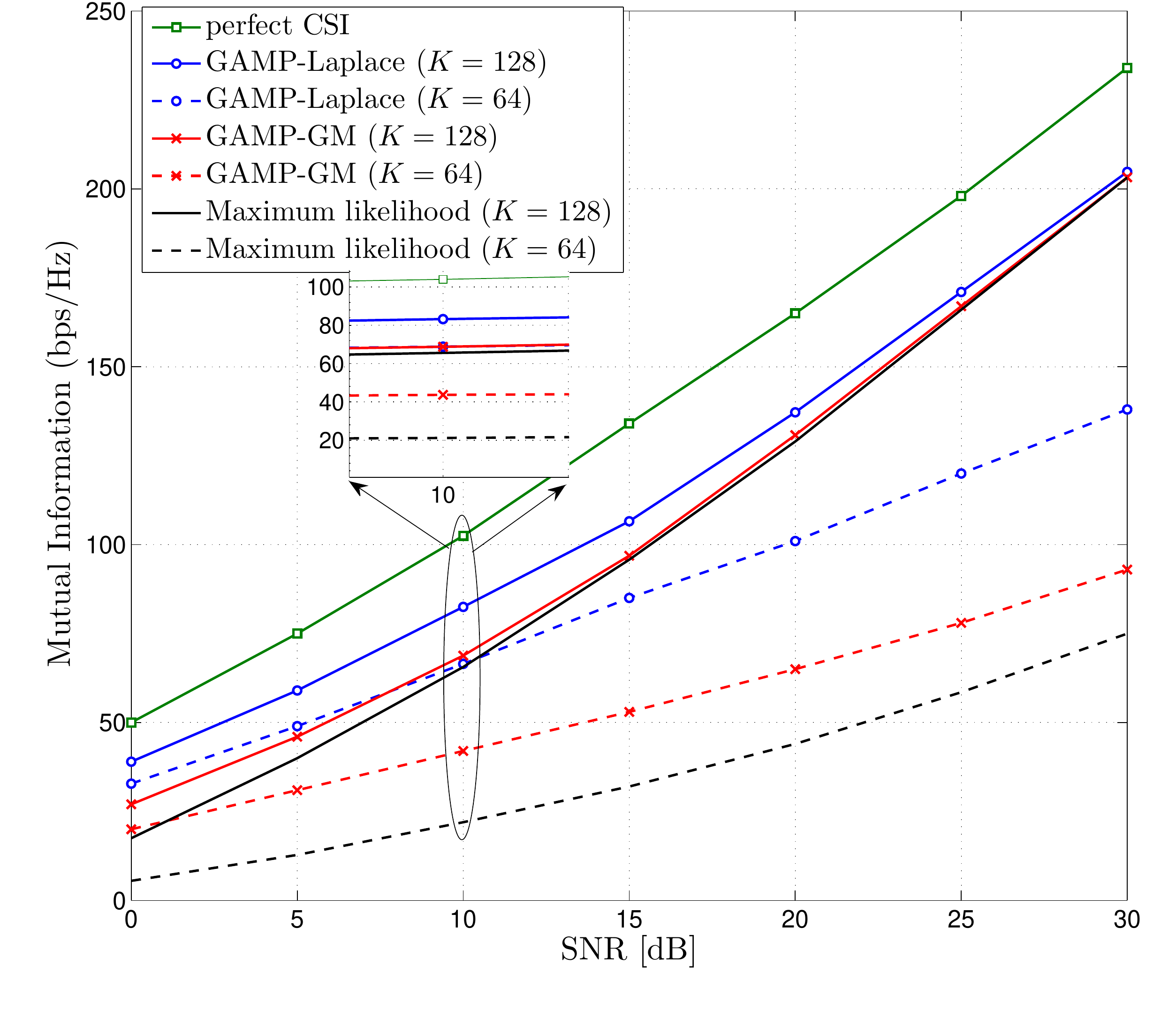}
\caption{Mutual information lower bound (\ref{lower_bound}) for ML, GAMP-GM, and GAMP-Laplace channel estimators versus SNR for training  window sizes $K=64$ and $K=128$.} 
\label{mutual_information}
\end{centering}
\end{figure}

The received signal, $\mathbf{y} = \mathbf{H}\mathbf{s}+\mathbf{w}$, is also pre-processed by $\widehat{\mathbf{U}}$. Under perfect CSI, this yields $S$ independent  parallel channels which can be decoded separately without loss of optimality. Channel estimation errors, however, introduce inter-stream interference which is independent of the background noise. By treating such interference as one more additive noise term while still decoding the streams separately, the mutual information between $\mathbf{s}$ and $\mathbf{y}$ given $\mathbf{H}$ is lower bounded\footnote{This is indeed a lower bound since it assumes a worst-case distribution for the interference components which is the Gaussian distribution.} by \cite{sparse_way}:
\begin{eqnarray}\label{lower_bound}
I'\big(\mathbf{s},\mathbf{y}\big|\mathbf{H}\big)&=&\sum_{m=1}^{M^*}\textrm{log}_2\left(1+\textrm{SINR}_m\right),
\end{eqnarray}
where $\textrm{SINR}_m$ is the signal-to-noise-plus-interference ratio (SINR) pertaining to the $m$th data streams:
\begin{eqnarray}
\!\!\!\!\!\!\!\!\!\!\!\textrm{SINR}_m&=&\frac{P_m\big|\widehat{\mathbf{u}}_m^{\textsf{H}}\,\mathbf{H}\,\widehat{\mathbf{v}}_m\big|^2}{2\sigma_w^2~+~\sum_{m'\neq m}P_{m'}\big|\widehat{\mathbf{u}}_{m}^{\textsf{H}}\,\mathbf{H}\,\widehat{\mathbf{v}}_{m'}\big|^2}.
\end{eqnarray}
Note here that for convenience we incorporate both active and non-active streams in (\ref{lower_bound}) since the latter do not change the final result as their  allocated powers are equal to zero (cf. Fig. \ref{waterfilling_figure}).

Fig. \ref{mutual_information} depicts the lower-bound mutual information (\ref{lower_bound}) achieved by the three estimators for two different training window sizes (namely, $K=64$ and $K=128$) versus SNR. The mutual information for the perfect CSI case is also plotted as an overall benchmark. In Fig. \ref{mutual_information}, we observe that GAMP-Laplace offers  remarkable throughput gains over the entire SNR range for $K=64$. In addition, although both GAMP-GM and the ML method perform nearly the same as GMAP-Laplace for very high SNR thresholds (when $K=128$), the latter is still quite advantageous for low-to-moderate SNRs. For instance, it offers  almost 15 bps/Hz throughput gain at $\textrm{SNR}=10$ dB as seen from  Fig. \ref{mutual_information}.

\section{Conclusion}\label{section_6}
In this paper, we propose a new channel estimator for massive MIMO mmWave systems that leverages the inherent sparsity of the channel in the angular domain. The proposed estimator belongs to the family of Bayesian estimators and builds upon the  generalized approximate message-passing  algorithm. This paper differs from most of the prior literature in that the angular-domain channel coefficients are modeled by a Laplacian prior distribution. We also propose a simple approach based on the expectation-maximization principle to systematically learn  the unknown scale parameter of the underlying Laplace distribution along with the unknown noise variance. 
 It is shown that a Laplacian prior leads to substantial performance improvements both in terms of channel estimation accuracy and achievable rate over the Gaussian mixture prior that has been advocated in the recent literature. Moreover, the Laplacian prior speeds up the convergence of GAMP thereby leading to significant computational savings in practice. As possible future directions, it will be interesting to investigate the correlation between the angular-domain channel coefficients and to incorporate the correlation into the estimation process and further to generalize the proposed algorithm to the multi-user case and to the case with finite-resolution analog-to-digital converters.            
 
 \section* {Appendix A}
  Using the Laplacian distribution, $p_{\mathcal{X}}(x_n;b)$, given in (\ref{Laplace_distribution}) and the fact that $|x| = \sgn(x)\,x,~\forall\,x\in\mathbb{R}$, it follows that:
\begin{eqnarray}\label{AppendixA_00}
\!\!\!\!\!\!\!\!\!p_{\mathcal{X}}(x_n;b)\mathcal{N}(x,\widehat{r}_n,\mu^r_n)&\!\!\!\!=\!\!\!\!& \frac{1}{2b}\frac{1}{\sqrt{2\pi\mu^r_n}}\exp\big(\!-\varphi(x_n)\big),
\end{eqnarray} 
where 
\begin{eqnarray}
\varphi(x_n) &\!\!\triangleq\!\!& \frac{(x_n^2-\widehat{r}_n)^2}{2\mu^r_n}~+~\frac{\sgn(x_n)x_n}{b},\nonumber\\
\!\!\!\!\!\!&\!\!=\!\!& \frac{1}{2\mu^r_n}\left[x_n^2\,-\,2\left(\widehat{r}_n-\frac{\sgn(x_n)\mu^r_n}{b}\right)x_n\,+\,\widehat{r}_n^2\right]\!.\nonumber
\end{eqnarray}
In order to complete the square inside the brackets, we add and subtract $\left[\widehat{r}_n-\sgn(x_n)\mu^r_n/b\right]^2$ thereby leading to:
\begin{eqnarray}
\varphi(x_n)&\!\!\!\!=\!\!\!\!& \frac{1}{2\mu^r_n}\Bigg[\left(x_n-\widehat{r}_n+\frac{\sgn(x_n)\mu^r_n}{b}\right)^2\nonumber\\
&&~~~~~~~~~~~~~~~~~~+\,\frac{\sgn(x_n)\widehat{r}_n}{b}2\mu^r_n\,-\,\left(\mu^r_n/b\right)^2\Bigg]\!.\nonumber
\end{eqnarray}
Then, by recalling the $(\widehat{r}_n,\mu^r_n,b)-$dependent functions $\alpha_n(x)$ and $\gamma_n(x)$ defined, respectively, in (\ref{alpha_function}) and (\ref{gamma_function}), it follows that:
\begin{eqnarray}\label{AppendixA_01}
\!\!\!\!\!\!\varphi(x_n) &\triangleq& \frac{(x_n^2-\gamma_n(x_n))^2}{2\mu^r_n}~+~\alpha_n(x_n).
\end{eqnarray}
Finally,  plugging (\ref{AppendixA_01}) back into (\ref{AppendixA_00}) yields:
\begin{eqnarray}\label{AppendixA_02} 
p_{\mathcal{X}}(x_n;b)\mathcal{N}(x,\widehat{r}_n,\mu^r_n)&&\nonumber\\
&& \!\!\!\!\!\!\!\!\!\!\!\!\!\!\!\!\!\!\!\!\!\!\!\!\!\!\!\!\!\!\!\!\!\!\!\!\!\!\!\!\!\!\!=\,\frac{\mathlarger{e^{-\alpha_n(x_n)}}}{2b}\frac{1}{\sqrt{2\pi\mu^r_n}}\exp\left\{\!-\frac{\Big(x_n-\gamma_n(x_n)\Big)^2}{2\mu^r_n}\right\}\!,\nonumber
\end{eqnarray} 
which is equivalent to  the result given claimed in (\ref{posterior_numerator}). 

\section* {Appendix B}
We have
\begin{eqnarray}
\!\!\!\!\!\!\!\!\!\sigma_{\mathcal{X}_n}^2(t+1)&& \nonumber\\
&&\!\!\!\!\!\!\!\!\!\!\!\!\!\!\!\!\!\!\!\!\!\!\!\!\!\!\!\!=\,\frac{1}{2b\psi_n}\!\int_{\mathbb{R}}\!x_n^2\mathlarger{e^{-\alpha_n(x_n)}}\mathcal{N}\big(x_n;\!~\gamma_n(x_n),\!~\mu^r_n\big)\,\mathrm{d}x_n.
\end{eqnarray}
By splitting the above integral into the positive and negative parts, it can be shown that:
\begin{eqnarray}\label{AppendixA_5}
\!\!\!\!\!\!\!\!\!\!\!\sigma_{\mathcal{X}_n}^2(t+1)&& \nonumber\\
&&\!\!\!\!\!\!\!\!\!\!\!\!\!\!\!\!\!\!\!\!\!\!\!\!\!\!\!\!\!\!\!\!= \frac{1}{2b\psi_n}\!\bigg[\mathlarger{e}^{-\alpha_n^+}\Phi_2\big(\gamma_n^+,\mu^r_n\big) + \mathlarger{e}^{-\alpha_n^-}\Phi_2\big(\!-\gamma_n^-,\mu^r_n\big) \bigg]\!,
\end{eqnarray}
in which the function $\Phi_2(\gamma,\mu)$ is defined as follows:
\begin{eqnarray}
\Phi_2\big(\gamma,\mu\big)&\triangleq& \frac{1}{\sqrt{2\pi\mu}}\int_{0}^{+\infty}t^2\mathlarger{e^{-\frac{(t-\gamma)^2}{2\mu}}}\,\mathrm{d}t.
\end{eqnarray}
Now, define:
\begin{eqnarray}
g(t)&=& \mathlarger{e}^{-\frac{(t-\gamma)^2}{2\mu}}~~~\longrightarrow ~~~~g'(t)~=~-\frac{t-\gamma}{\mu}g(t)\\
f(t) &=& t~~~~~~~~~~~~~\!\longrightarrow ~~~~f'(t)~=~1
\end{eqnarray}
Using integration by parts, it follows that:
\begin{eqnarray}
\int_{0}^{+\infty}f(t)g'(t)\,\mathrm{d}t&=& \big[f(t)g(t)\big]_0^{+\infty}~-~\int_{0}^{+\infty}f'(t)g(t)\,\mathrm{d}t.\nonumber\\
\end{eqnarray}
Since $\big[f(t)g(t)\big]_0^{+\infty}=0$,  we have:
\begin{eqnarray}\label{eq.1}
\int_{0}^{+\infty}\frac{t(t-\gamma)}{\mu}\mathlarger{e}^{-\frac{(t-\gamma)^2}{2\mu}}\,\mathrm{d}t&=& \int_{0}^{+\infty}\mathlarger{e}^{-\frac{(t-\gamma)^2}{2\mu}}\,\mathrm{d}t,
\end{eqnarray} 
Then, using the substitution $x=(t-\gamma)/\sqrt{\mu}$ in the right-hand side of (\ref{eq.1}) and expanding its left-hand side leads to:
\begin{eqnarray}\label{eq.2}
\frac{\sqrt{2\pi\mu}}{\mu}\Phi_2\big(\gamma,\mu\big) - \frac{\gamma\sqrt{2\pi\mu}}{\mu}\Phi_1\big(\gamma,\mu\big)&=& \sqrt{\mu}\int_{-\frac{\gamma}{\sqrt{\mu}}}^{+\infty}\mathlarger{e}^{-\frac{x^2}{2}}\,\mathrm{d}t\nonumber\\
&=&\sqrt{2\pi\mu}\!~\mathsf{Q}(-\gamma/\sqrt{\mu}),\nonumber\\
\end{eqnarray}
Finally, after rerranging the terms in (\ref{eq.2}), it follows that:
\begin{eqnarray}\label{AppendixA_6} 
\Phi_2\big(\gamma,\mu\big) &=&\gamma\Phi_1\big(\gamma,\mu\big)~+~\mu\textsf{Q}\big(\!-\gamma/\sqrt{\mu}\big).
\end{eqnarray}
By recalling (\ref{AppendixA_2}) and using the identity in (\ref{Appendix_A_identity}), it can be shown that: 
\begin{eqnarray}\label{AppendixA_7} 
\!\!\!\!\!\!\!\!\!\!\mathlarger{e}^{-\alpha_n^+}\Phi_2\big(\gamma_n^+,\mu^r_n\big)&&\nonumber\\
&&\!\!\!\!\!\!\!\!\!\!\!\!\!\!\!\!\!\!\!\!\!\!\!\!\!\!\!\!\!\!\!\!\!\!\!\!\!\!\!\!\!\!\!\!\!\!\!\!\!\!=\Big((\gamma_n^+)^2+\mu^r_n\Big) \mathlarger{e}^{-\alpha_n^+}\textsf{Q}\left(\!\frac{-\gamma_n^+}{\sqrt{\mu^r_n}}\right) + \frac{\mu^r_n\gamma_n^+}{\sqrt{2\pi\mu^r_n}}\mathlarger{e^{-\frac{\widehat{r}_n^2}{2\mu_n^r}}},\\
\label{AppendixA_8}\!\!\!\!\!\!\!\!\!\!\mathlarger{e}^{-\alpha_n^-}\Phi_2\big(\!\!-\gamma_n^-,\mu^r_n\big)&&\nonumber\\
&&\!\!\!\!\!\!\!\!\!\!\!\!\!\!\!\!\!\!\!\!\!\!\!\!\!\!\!\!\!\!\!\!\!\!\!\!\!\!\!\!\!\!\!\!\!\!\!\!\!\!=\Big(\!(\gamma_n^-)^2+\mu^r_n\Big) \mathlarger{e}^{-\alpha_n^-}\textsf{Q}\left(\!\frac{\gamma_n^-}{\sqrt{\mu^r_n}}\right) - \frac{\mu^r_n\gamma_n^-}{\sqrt{2\pi\mu^r_n}}\mathlarger{e^{-\frac{\widehat{r}_n^2}{2\mu_n^r}}}.
\end{eqnarray} 
Now,  plugging (\ref{AppendixA_7}) and  (\ref{AppendixA_8}) back in (\ref{AppendixA_5}) and using the fact that:  
\begin{eqnarray}
\gamma_n^+~-~\gamma_n^-&=& -\frac{2\mu_n^r}{b},   
\end{eqnarray}
yields the result claimed in (\ref{AppendixA_9}). 
 
\bibliographystyle{IEEEtran}
\bibliography{IEEEabrv,refrences} 
\begin{IEEEbiography}[{\includegraphics[width=1in,height=1.25in,clip,keepaspectratio]{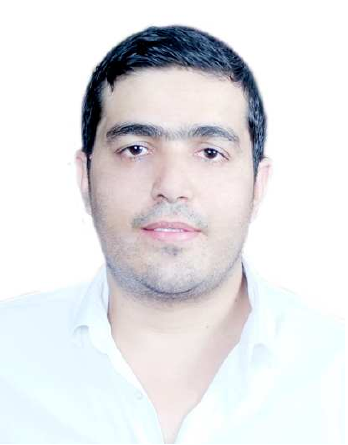}}]{Faouzi Bellili} (M'19)
\changer{received the B.Eng. degree (Hons.) in electrical engineering from Tunisia Polytechnic School, in 2007, and the M.Sc. degree (Hons.) and Ph.D. degree (Hons.) from the National Institute of Scientific Research (INRS), University of Quebec, Montreal, QC, Canada, in 2009 and 2014, respectively. From 2014 to 2016, he was a Research Associate with INRS-EMT, where he coordinated a major multi-institutional NSERC Collaborative R\&D (CRD) Project on 5th--Generation (5G) Wireless Access Virtualization Enabling Schemes. From 2016 to 2018, he was a Post-Doctoral Fellow with the Electrical and Computer Engineering Department, University of Toronto, Toronto, ON, Canada. He is currently an Assistant Professor with the Department of Electrical and Computer Engineering, University of Manitoba, Winnipeg, MB, Canada. His research focuses on statistical and array signal processing for wireless communications and 5G-enabling technologies. He serves regularly as a TPC member for the major IEEE conferences. He received the very prestigious NSERC PDF Grant (2017--2018), and also a prestigious PDF Scholarship offered over the same period (but declined) from the Fonds de Recherche du Quebec Nature et Technologies. He was a recipient of the INRS Innovation Award in 2015, the very prestigious Academic Gold Medal of the Governor General of Canada (2009--2010), and the Excellence Grant of the Director General of INRS (2009--2010). He received the award for the best M.Sc. Thesis at INRS-EMT (2009--2010), and twice---for both the M.Sc. and Ph.D.
programs---the National Grant of Excellence from the Tunisian Government. In 2011, he received the Merit Scholarship for Foreign Students from the Ministere de l'Education, du Loisir et du Sport of Quebec, Canada. He acts as a reviewer for many international scientific journals and conferences.}

\end{IEEEbiography} 

\begin{IEEEbiography}[{\includegraphics[width=1in,height=1.25in,clip,keepaspectratio]{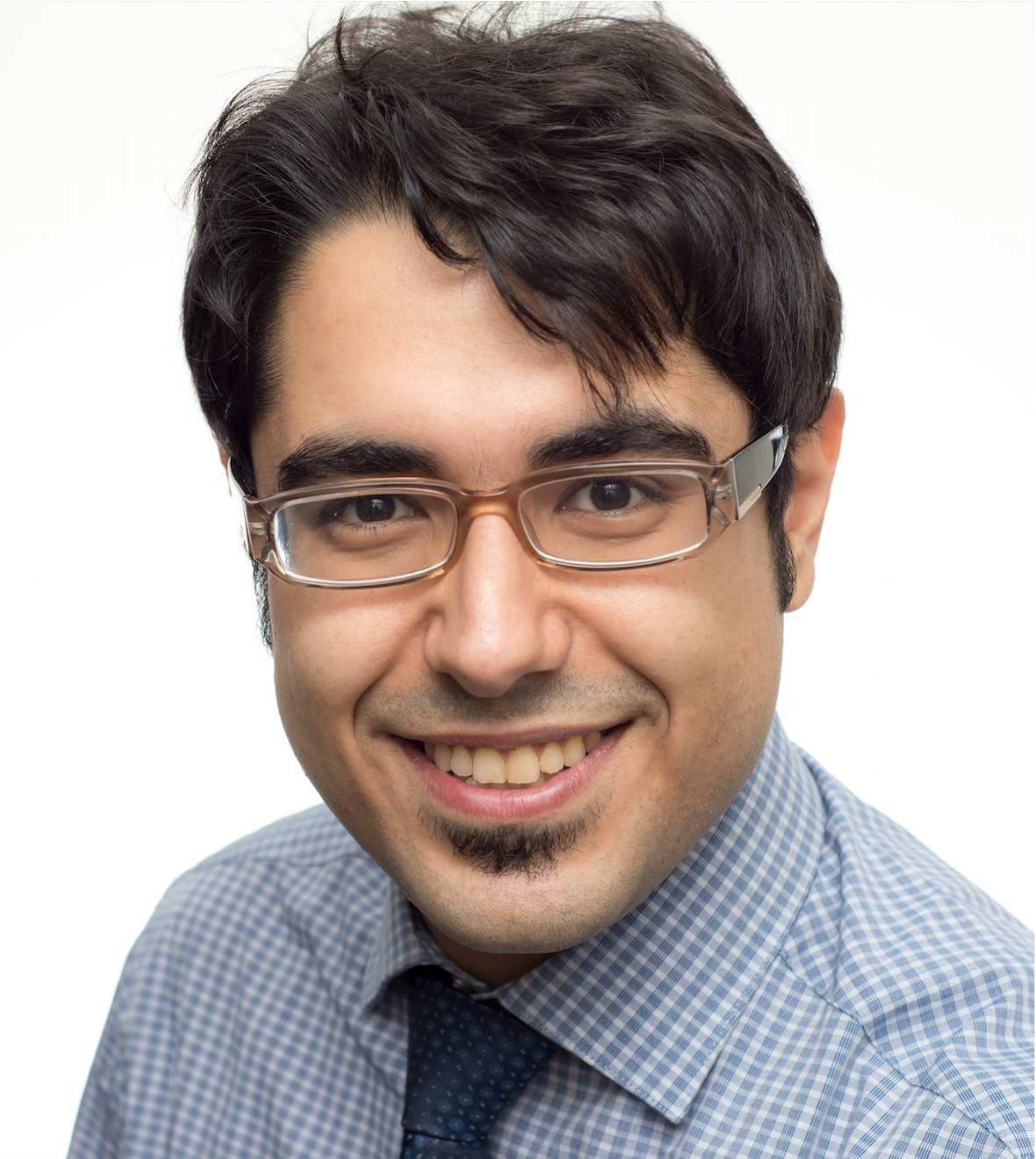}}]{Foad Sohrabi}
  (S'13--M'19) received the B.A.Sc. degree from University of Tehran, Tehran, Iran, in 2011, the M.A.Sc. degree from McMaster University, Hamilton, ON, Canada, in 2013, and the Ph.D. degree from University of Toronto, Toronto, ON, Canada, in 2018,  all in electrical and computer engineering. Since \changeb{2018}, he has been \changeb{a Post-Doctoral} Fellow \changeb{with the} University of Toronto. \changeb{In 2015,} he was a Research Intern with Bell Labs, Alcatel-Lucent, Stuttgart, Germany. His research interests include MIMO communications, optimization theory, wireless communications, signal processing and machine learning. He was \changeb{a} recipient of the IEEE Signal Processing Society Best Paper Award in 2017.
\end{IEEEbiography}

\begin{IEEEbiography}[{\includegraphics[width=1in,height=1.25in,clip,keepaspectratio]{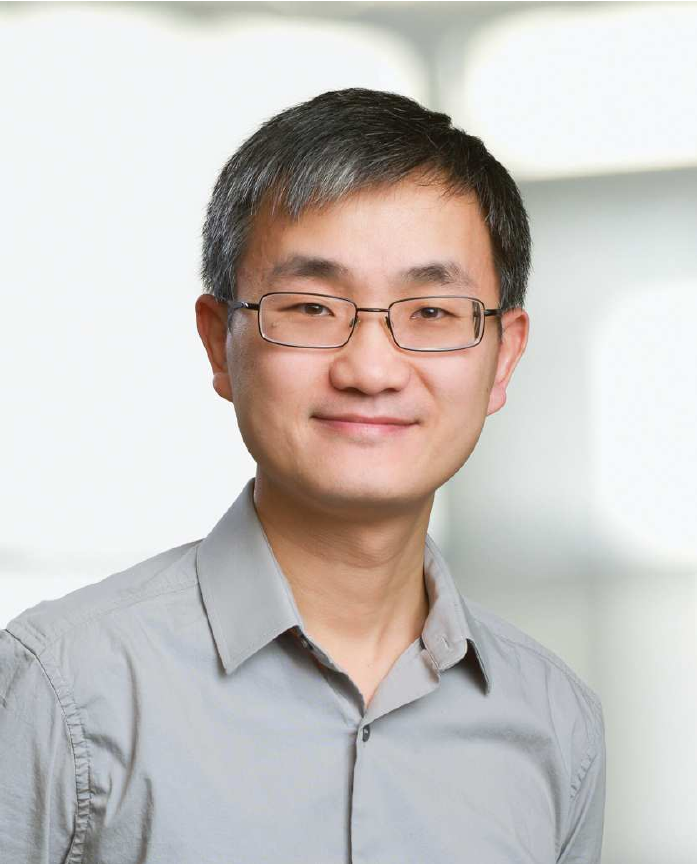}}]{Wei Yu}
   (S'97--M'02--SM'08--F'14) received the B.A.Sc. degree in \changer{computer
engineering and mathematics} from the University of Waterloo, Waterloo,
\changer{ON, Canada, in 1997,} and \changer{the M.S.} and Ph.D. degrees in \changer{electrical
engineering} from Stanford University, Stanford, CA, \changer{USA,} in 1998 and 2002,
respectively. Since 2002, he has been with the Electrical and Computer
Engineering Department, \changer{University of Toronto,} Toronto, \changer{ON},
Canada, where he is \changer{currently a} Professor and holds a Canada Research Chair
(Tier 1) in \changer{information theory and wireless communications.} His main
research interests include information theory, optimization, wireless
communications, and broadband access networks.

\changer{Prof. Yu is a Fellow of the Canadian Academy of Engineering, and a member of the College of New Scholars, Artists and Scientists of the Royal Society of Canada. He serves as the Second Vice President of the IEEE Information Theory Society in 2019. He received the Steacie Memorial Fellowship in 2015, the IEEE Signal Processing Society Best Paper Award 
in 2017 and 2008, respectively, the \textit{Journal of Communications and Networks} Best Paper Award in 2017, the IEEE Communications Society Best Tutorial Paper Award in 2015, the IEEE ICC Best Paper Award in 2013, the McCharles Prize for Early Career Research Distinction in 2008, the Early Career Teaching Award from the Faculty of Applied Science and Engineering, University of Toronto, in 2007, and an Early Researcher Award from Ontario, in 2006. He has served as the Chair of the Signal Processing for Communications and Networking Technical Committee of the IEEE Signal Processing Society (2017--2018), and as a member (2008--2013). He has served as an Associate Editor for the IEEE TRANSACTIONS ON INFORMATION THEORY (2010--2013), as an Editor for the IEEE TRANSACTIONS ON COMMUNICATIONS (2009--2011), and as an Editor for the IEEE TRANSACTIONS ON WIRELESS COMMUNICATIONS (2004--2007). He is currently an Area Editor of the IEEE TRANSACTIONS ON WIRELESS COMMUNICATIONS (2017--2020). He is recognized as a Highly Cited Researcher. He was an IEEE Communications Society Distinguished Lecturer (2015--2016).

}

\end{IEEEbiography}

\end{document}